\documentclass[12pt]{article}
\usepackage{latexsym}
\usepackage[dvips]{graphics,color}
\usepackage{graphpap}
\usepackage{alltt}
\usepackage{graphicx}
\usepackage{amsfonts}
\usepackage{psfrag}

\def\pr#1#2{Phys. Rep. {\bf{#1}}, #2}

\def\pre#1#2{Phys. Rev. {\bf{E#1}}, #2}
\def\prl#1#2{Phys. Rev. Lett. {\bf{#1}}, #2}

\def\rmp#1#2{Rev. Mod. Phys. {\bf{#1}}, #2}

\def\apk#1#2{Ann. Physik {\bf{#1}}, #2}

\def\jsp#1#2{J. Stat. Phys. {\bf{#1}}, #2}

\def\fcaa#1#2{Fractional Calculus and Applied Analysis {\bf{#1}}, #2}
\def\jcam#1#2{J. of Computational and Applied Mathematics {\bf{#1}}, #2}

\def\jmp#1#2{J. Math. Phys. {\bf{#1}}, #2}

\def\rmp#1#2{Rev. Mod. Phys. {\bf{A#1}}, #2}
\def\tams#1#2{Trans. Amer. Math. Soc. {\bf{A#1}}, #2}

\def\apt#1#2{Acta Protozool. {\bf{#1}}, #2}

\renewcommand{\baselinestretch}{1}

\begin{document}

\begin{center}
{\Large \bf One dimensional stable probability density functions for rational index $\bf 0<\alpha \leq 2$} \\ [4mm]
\large{AGAPITOS HATZINIKITAS} \\ [5mm]
{\small
University of Aegean, \\
School of Sciences,\\
Department of Mathematics, \\
83200 Karlovasi, Samos Greece.  \\
Email: ahatz@aegean.gr }  \\ [5mm]

\large{and} \\ [5mm]

\large{JIANNIS K. PACHOS } \\ [5mm]
{\small School of Physics and Astronomy, \\
University of Leeds,
Leeds LS2 9JT, U.K.\\
Email: j.k.pachos@leeds.ac.uk}

\end{center}

\begin{abstract}
Fox's H-function provide a unified and elegant framework to tackle several
physical phenomena. We solve the space fractional diffusion equation on the
real line equipped with a delta distribution initial condition and identify
the corresponding H-function by studying the small $x$ expansion of the
solution. The asymptotic expansions near zero and infinity are expressed,
for rational values of the index $\alpha$, in terms of a finite series of
generalized hypergeometric functions. In $x$-space, the $\alpha=1$ stable
law is also derived by solving the anomalous diffusion equation with an
appropriately chosen infinitesimal generator for time translations. We
propose a new classification scheme of stable laws according to which a
stable law is now characterized by a generating probability density
function. Knowing this elementary probability density function and bearing
in mind the infinitely divisible property we can reconstruct the
corresponding stable law. Finally, using the asymptotic behavior of
H-function in terms of hypergeometric functions we can compute closed
expressions for the probability density functions depending on their
parameters $\alpha, \beta, c, \tau $. Known cases are then reproduced and
new probability density functions are presented.
\end{abstract}

\noindent\textit{Key words:} Probability distributions, L\'{e}vy flights, Integral equations, Special functions.\\
\textit{PACS:} 02.50.-r, 05.40-a, 02.30.Rz, 02.30.Gp

\section{Introduction}
\label{intro}

\par The notion of stable distributions was first introduced by L\'evy \cite{Ref1} in the study of Generalized Central Limit Theorem, and there is a nice early account of the theory in \cite{Ref2}. A stable law is a direct generalization of the Gaussian distribution and in fact includes the Gaussian as a limiting case. The main difference between the stable and the Gaussian distributions is that the tails of the stable density are heavier than those of the Gaussian density. This characteristic is one of the main reasons why stable laws are suitable for modelling a plethora of phenomena such as laser cooling \cite{Ref3}, turbulence \cite{Ref4}, dynamical systems \cite{Ref5}, statistical mechanics, signal processing \cite{Ref6}, biology \cite{Ref7} and mathematical finance
\cite{Ref8}.
\par The goal of the present paper is to give analytic expressions for the probability density functions (p.d.f.'s) of the stable laws, in terms of known functions. These were lacking from the literature apart from a handful of well known ones. To accomplish our task we organize the paper as
follows:
\par In \textit{Section 2} we briefly review the definition of infinite divisible laws in terms of characteristic functions (or the Fourier transform of the probability measure) and limit our investigation to the subclass of stable laws. We give the characteristic exponent for the general one-dimensional case and comment on the role and the essential properties of the parameters which are
involved.
\par In \textit{Section 3} we exploit the definition of the Fox's $H$-function as a Mellin-Barnes path integral and performing the integration on the appropriate contour. This allows us to write its asymptotic expansion at zero and infinity
(see expressions (\ref{fh4}) and (\ref{fh5})). The reason for choosing the
specific values $m=n=1$ and $p=q=2$ for the $H$-function is its close
relation to the solution of the free space fractional diffusion equation.
Using the Gauss's multiplication formula as well as standard properties of
the gamma function we resum the series for rational values of the index
$\alpha$ and produce general closed expressions containing the generalized
hypergeometric functions. These expressions can be manipulated, for
different values of the parameters, using a simple computer program running
under Maple software.
\par In \textit{Section 4} we solve the anomalous spatial diffusion equation on the real line with a fractional Laplacian consisting of Weyl derivatives and a Dirac delta distribution as initial condition.
In this way we determine the most general form of the p.d.f. for a stable
law. We establish the connection with the corresponding Fox function and
find its asymptotics. Finally, we express them as finite sums of generalized
hypergeometric functions. The fundamental solution (Green function) for the Cauchy problem of the space-time fractional diffusion equation as well as its relation to the Mejer G-functions have been studied by \cite{Ref9, Ref10}.
\par In \textit{Section 5} we demonstrate that the diffusion equation has
as infinitesimal generator of time translations the spatial derivative of
the convolution of two terms, which under Fourier transformation they
reproduce the characteristic function of the $\alpha=1$ stable law. The
corresponding integral cannot be performed exactly, unless $\beta=0$ in
which case we recover the shifted Cauchy p.d.f. Only in the small $x, \beta$
regime one can provide a triple series expansion.
\par In \textit{Section 6} we establish a new way to classify stable laws
by exploiting their infinitely divisible property. It is possible to write a
formula that determines the p.f.d. of the stable law as an infinite limit of
the m-fold convolution of a generating p.d.f. This expression although it
gives new insight, from calculational view point is cumbersome due to its
complexity.
\par In \textit{Section 7} we present a sample of our results in the
subdiffusion regime ($\alpha<1$) while $\alpha$ takes values on the Farey
series \footnote{The Farey series $\mathcal{F}_n$ of order n is the
ascending series of irreducible fractions between $0$ and $1$ whose
denominators do not exceed $n$. Thus $\alpha=\frac{p}{q}$ belongs in
$\mathcal{F}_n$ if
\begin{displaymath}0\leq p \leq q \leq n ,
\quad (p,q)=1 \end{displaymath}
where $(,)$ denotes the highest common divisor of two integers.}
$\mathcal{F}_n$ of order $n=5$. In the superdiffusion regime ($\alpha>1$) we
recover all previously known results and also give new ones for general
rational $\alpha$.
\section{Preliminaries on $\bf{\alpha}$-stable laws}
\label{sec:1}

Consider a probability measure $\mu$ on ${\mathbb R}^n$ and its characteristic function
\cite{Ref11,Ref12,Ref13}
\begin{eqnarray}
\hat{\mu}(p)=\mathcal{F}[\mu](p)=\int_{{\mathbb R}^n}e^{i<p,x>}\mu(dx), \,\, p\in {\mathbb R}^n.
\label{ps1}
\end{eqnarray}
\noindent A probability measure $\mu$ on ${\mathbb R}^n$ is called \textit{infinitely divisible} if
\begin{displaymath}
\forall m\in N, \,\, \exists \,\, \mu_m, \,\, \mathcal{F}[\mu_m] \, :
\mathcal{F}[\mu](p)=\left(\mathcal{F}[\mu_m](p)\right)^m
\end{displaymath}
where $\mu=\mu_m*\cdots *\mu_m$ is the m-fold convolution of $\mu_m$ with itself.
If the measure $\mu$ is infinitely divisible then there exists a unique continuous function
$\psi : {\mathbb R}^n\rightarrow {\mathbb C}$, called the characteristic exponent of $\mu$, such that $\psi(0)=0$
and
\begin{displaymath}
\mathcal{F}[\mu](p)=e^{\Psi(p)}, \,\, p\in {\mathbb R}^n.
\end{displaymath}
The \textit{L\'{e}vy-Khintchine
representation} or \textit{L\'{e}vy-Khintchine formula} states that a
probability measure $\mu$ on ${\mathbb R}^n$ is infinitely divisible iff we can write the
characteristic exponent in the form
\begin{eqnarray}
\psi(p)=i<p,\tau>-\frac{1}{2}<p,Ap>+\int_{{\mathbb R}^n}\left(e^{i<p,x>}-1-i<p,x>
1_{|x|\leq 1}\right) \nu(dx)
\label{ps2}
\end{eqnarray}
\noindent where $\tau\in {\mathbb R}^n$, A is a symmetric nonnegative-definite $n\times n$
matrix, called the \textit{Gaussian covariance matrix}, and $\nu$ is a
$\sigma$-finite Borel measure on ${\mathbb R}^n_0:={\mathbb R}^n / \{0\}$, called the
\textit{L\'{e}vy measure}, such that
\begin{eqnarray}
\int_{{\mathbb R}_0^n}{\rm min} \{1,||x||^2 \} \nu(dx)<\infty.
\label{ps3}
\end{eqnarray}
\noindent The triplet $[\tau,A,\nu]$ is unique
and will be called the generating triplet of the infinitely divisible
probability measure $\mu$. If $A=0$ then $\mu$ is said to be purely
non-Gaussian.
\par A subclass of infinitely divisible laws is the stable laws class. Suppose that
$\textbf{X},\textbf{X}_1, \cdots , \textbf{X}_m$ denote mutually independent random variables
with a common distribution F and $\textbf{S}_m=\sum_{i=1}^{m}\textbf{X}_i$.
The distribution F is stable if for each $m\in {\mathbb N}$ there exist constants $c_m>0$ and
$\tau_m \in {\mathbb R}$ such that
\begin{eqnarray}
\textbf{S}_m\stackrel{d}{=}c_m \textbf{X}+\tau_m
\label{ps4}
\end{eqnarray}
\noindent and F is not concentrated at one point. F is stable in the strict sense if $\tau_m=0$.
The symbol $\stackrel{d}{=}$ means that the distributions of $\textbf{S}_m$ and
$\textbf{X}$ are identical up to scale and location parameters. The norming constants are of the form $c_m=m^{\frac{1}{\alpha}}$ with
$0<\alpha\leq 2$ and the constant $\alpha$ is called \textit{characteristic exponent} of F or
\textit{index} of the stable law.
\par Let $0<\alpha<2$ and $\mu$ be an infinitely divisible and non-trivial on ${\mathbb R}^n$ probability measure with generating triplet $[\tau,A,\nu]$. If $\mu$ is $\alpha$-stable then there is a finite non-zero measure $\lambda$ on the unit sphere $S=\{x\in {\mathbb R}^n: \, |x|=1\}$ such that
\begin{description}
\item[($i$)] $A=0$ and $\nu(B)=\int_S\lambda(d\xi)\int_{0}^{\infty}1_B(r\xi)\frac{d\xi}{r^{1+\alpha}}$
for $B\in\mathcal{B}({\mathbb R}^n)$.
\item[($ii$)] $\hat{\mu}(p)=exp\left[-\int_S |<p,\xi>|^{\alpha}(1-i\tan(\frac{\pi \alpha}{2}))\textrm{sgn}<p,\xi>
)\lambda(d\xi)+i<\tau,p>\right]$ for $\alpha\neq 1$ and $\tau\in {\mathbb R}^n$
\item[($iii$)] $\hat{\mu}(p)=exp\left[-\int_S (|<p,\xi>| +i\frac{2}{\pi}<p,\xi>
\ln|<p,\xi>|)\lambda(d\xi)+i<\tau,p>\right]$ for $\alpha=1$ and $\tau\in {\mathbb R}^n$.
\end{description}
In the one-dimensional case ($n=1$) one can prove that the characteristic exponent has the form
\begin{eqnarray}
\psi(p)=i\tau p-\Biggl\{
\begin{array}{cl} c|p|^{\alpha}\left[1-i\beta \, \textrm{sgn}(p)\tan (\frac{\pi \alpha}{2}) \right]
& \mbox{if} \quad \alpha \neq 1, 2 \\
c |p|\left[1+i \beta \frac{2}{\pi}\, \textrm{sgn}(p)\ln (|p|)\right]& \mbox{if} \quad \alpha=1 \end{array}
\label{ps7}
\end{eqnarray}
\noindent where
\begin{eqnarray}
\beta= \frac{c_{+} -c_{-}}{c_{+} +c_{-}}, \quad \textrm{and} \quad c=\frac{\pi}{2\Gamma(1+\alpha)}\frac{1}{\sin \left(\frac{\pi \alpha}{2}\right)}(c_++c_-), \quad \mbox{for} \quad \alpha\neq 2
\label{ps71}
\end{eqnarray}
\noindent with $c_+, c_- \geq 0$ and $c_++c_->0$ \footnote{The same result can be recovered if one uses the
absolutely continuous L$\acute{e}$vy measure
\begin{displaymath}
\nu(dx)=\left(c_+ 1_{x>0}+c_-
1_{x<0}\right)|x|^{-1-\alpha}dx.
\end{displaymath}
}. Note that when $\alpha=2$ then $\nu=0$.
\par The collection of the four parameters $(\alpha, \beta, c, \tau)$ is called the \textit{stable law parameters} and completely determines the distribution as follows: \\
\textit{Characteristic exponent $\alpha$}. This parameter determines the degree of leptokurtosis
and the fatness of the tails. For a stable real-valued random variable $X$ it can be shown that
\begin{displaymath}
E|X|<\infty \Longleftrightarrow 1<\alpha \leq 2.
\end{displaymath}
When $\alpha \leq 1$ the means becomes infinite. The variance for $\alpha \in (0,2)$
becomes infinite or undefined while all moments of a random variable X become finite
iff $\alpha=2$. Also the moments of order less than $\alpha$ are positive and
have a finite limit, namely $E|X|^k<\infty, \,\, 0<k<\alpha$ \footnote{It is also true that a symmetric $\alpha$-stable random variable has finite negative-order moments $-1<k<0$ \cite{Ref14}.}. \\
\textit{Skewness parameter $\beta$}. This parameter characterizes the degree of asymmetry
of the L$\acute{e}$vy measure and takes values in the interval $[-1,1]$. The measure
$\nu$ is called symmetric if $\beta=0$ (or $c_+=c_-$) and the
$\alpha$-stable distribution is called stable $\alpha$-symmetric
\footnote{In general a measure $\mu$ is symmetric when $\mu(B)=\mu(-B)$ for
$B\in \mathcal{B}({\mathbb R}^n)$. In $n=1$ the rotation invariance is tantamount to symmetry.}
.\\
\textit{Scale parameter c.} This parameter ranges into the interval $(0,\infty)$ and measures
scale in place of standard deviation. \\
\textit{Location parameter $\tau$.} This parameter saturates the set of real numbers
and shifts the distribution to the left or right. If $1<\alpha <2$ then $\tau$ equals to the
mean of $\mu$. When $0<\alpha<1$, although, the mean is infinite it serves as an
index of the location of the peak of the stable distribution and is identical to the drift of
$\mu$.
\par A stable law generated
by $(\alpha, \beta, c, \tau)$ is often denoted by $S_{\alpha}(\beta,c,\tau)$. In the
present work our law will be first generated by $S_{\alpha}(c)$ and then we will study the most general case.
\par The parameter space $(\alpha, b)$ of stable p.d.f.'s with the centering constant b restricted in the region \cite{Ref15}
\begin{eqnarray}
|b|\leq \Biggl\{ \begin{array}{ll} \alpha, & \quad 0<\alpha<1 \\ 2-\alpha, & \quad 1\leq \alpha \leq 2 \end{array}
\label{ps8}
\end{eqnarray}
is depicted in the following figure
\begin{figure}[!h]
\centering
\psfrag{1}[c]{\,\,\footnotesize{1}}
\psfrag{-1}[c]{\,\,\,\, \footnotesize{-1}}
\psfrag{1}[c]{\,\,\footnotesize{1}}
\psfrag{2}[c]{\footnotesize{2}}
\psfrag{1/2}{$\frac{1}{2}$}
\psfrag{3/2}{\,\,\,\, $\frac{3}{2}$}
\psfrag{x}[l]{\footnotesize{$\alpha$-parameter}}
\psfrag{y}[c][][1][90]{\footnotesize{b-parameter}}
\includegraphics[scale=1]{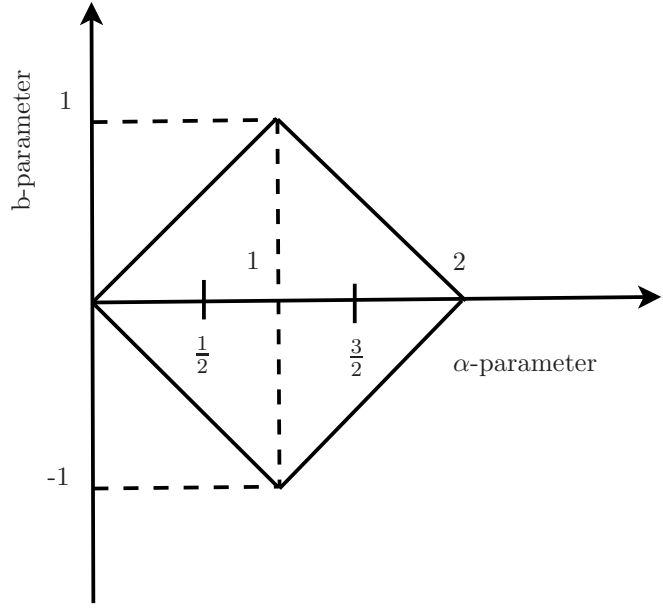}
\caption{The parameter space of all stable p.d.f's on the $\tau=0$ plane.
The points of the axis $\beta=0$ represent the stable $\alpha$-symmetric p.d.f.'s with the property
$f_{\alpha, \beta=0}(x)=f_{\alpha, \beta=0}(-x)$. On this axis are
located the familiar Cauchy ($\alpha=1$),
Holtsmark ($\alpha=\frac{3}{2}$) and Gaussian ($\alpha=2$) distributions.}
\label{fig:1}
\end{figure}
\section{Fox's $H$-function and generalized hypergeometric functions for the law  \boldmath \( S_{\alpha}(c=K_{\alpha}) \) with rational \boldmath \( \alpha \)}
\label{sec:2}

\par Fox \cite{Ref16,Ref17} defined the $H$-function in his studies of symmetrical Fourier kernels as the
Mellin-Barnes path integral
\begin{eqnarray}
H_{p, q}^{m, n}(z)=H_{p, q}^{m, n}\left[ z \Biggl|
\begin{array}{cl} (a_1, A_1), (a_2, A_2), \cdots, (a_p, A_p) \\
(b_1, B_1), (b_2, B_2), \cdots, (b_q, B_q)\end{array}\right]=\frac{1}{2\pi i}\int_{\mathcal{C}}
\chi(s)z^s ds
\label{fh1}
\end{eqnarray}
\noindent where the integral density $\chi(s)$ is given by
\begin{eqnarray}
\chi(s)=\frac{\prod_{i=1}^{m}\Gamma(b_{i}-B_{i}s)
\prod_{j=1}^{n}\Gamma(1-a_{j}+A_{j}s)}
{\prod_{i=m+1}^{q}\Gamma(1-b_{i}+B_{i}s)
\prod_{j=n+1}^{p}\Gamma(a_{j}-A_{j}s)},
\label{fh2}
\end{eqnarray}
\noindent $m, n, p, q$ are integers satisfying
\begin{displaymath}
0\leq n\leq p, \quad 0\leq m \leq q,
\end{displaymath}
$B_i, A_j$ are positive numbers and $b_i, a_j$ are complex numbers such that
\begin{displaymath}
A_j(b_h+\nu)\neq B_h(a_j-1-\lambda), \quad \nu, \lambda=0,1 \cdots; \quad h=1,\cdots,m;
\,\, j=1,\cdots,n.
\end{displaymath}
This condition implies that the poles of $\Gamma(b_{i}-B_{i}s)$ and
$\Gamma(1-a_{j}+A_{j}s)$ form two disjoint sets. $\mathcal{C}$ is a contour
in the complex $s$-plane which runs from $s=\infty-ik$ to $s=\infty+ik$ with
$k>\frac{|{\rm Im} b_j|}{B_j}, \,\,j=1,\cdots,n$ and which encloses the
poles
\begin{displaymath}
s=\frac{(b_i+\nu)}{B_i} \quad  i=1,\cdots,m
\end{displaymath}
but none of the poles
\begin{displaymath}
s=\frac{(a_j-1-\nu)}{A_j} \quad j=1,\cdots,n.
\end{displaymath}
$H(z)$ makes sense and defines an analytic function of $z$ in
the following two cases:
\begin{description}
\item[($i$)] If
\begin{eqnarray}
M=\sum_{i=1}^{q}B_i-\sum_{j=1}^{p}A_j>0, \quad \forall z\neq 0
\label{fh21}
\end{eqnarray}
\item[($ii$)] If
\begin{eqnarray}
M=0 \quad \textrm{and} \quad  0<|z|<R \quad \textrm{with}
\quad R=\prod_{i=1}^{q}(B_i)^{B_i} \prod_{j=1}^{p}(A_j)^{-A_j}.
\label{fh22}
\end{eqnarray}
\end{description}
\par Schneider \cite{Ref18} has introduced into physics the Fox's
$H$-function as analytic representations for the L$\acute{e}$vy
distributions in x-space, and as solutions of the fractional diffusion
equation \footnote{In nature there is a diversity of diffusion processes for
which the $k$-moment of the displacement grows not linearly with time but
follows a power-law pattern of the form $E|X(t)|^k \sim
t^{\frac{k}{\alpha}}, \,\, 0<k<\alpha$.}. If one tries to solve the
one-dimensional anomalous diffusion equation equipped with the following
initial and boundary conditions\footnote{In equation (\ref{fh2a})
$K_{\alpha}$ is the generalized diffusion constant having dimensions
$[L]^{\alpha} [T]^{-1}$ and $\hat{\mathcal{A}}$ is the operator
(\ref{3a3a2}) with $\beta=0$ and $\tau=0$. This is the generalization of the
Laplacian to a fractional order.}
\begin{eqnarray}
\frac{\partial f(x,t)}{\partial t}&=& K_{\alpha} \hat{\mathcal{A}}(\alpha) f(x,t), \quad x\in {\mathbb R}, \, t>0, \, \alpha\in(0,2)  \nonumber \\
\lim_{t\downarrow 0}f(x,t)&=&\delta(x), \quad \lim_{|x|\rightarrow \infty}f(x,t)=0
\label{fh2a}
\end{eqnarray}
then the solution (or propagator) reads
\begin{eqnarray}
f(x,t; \alpha)=\frac{1}{\alpha |x|}H_{2, 2}^{1, 1}
\left[ \frac{|x|}{(K_{\alpha}t)^{\frac{1}{\alpha}}}
\Biggl|
\begin{array} {cl}(1, \frac{1}{\alpha}), & (1, \frac{1}{2})\\
(1, 1), & (1, \frac{1}{2}) \end{array}\right]
\label{fh2b}
\end{eqnarray}
which expresses the $\alpha$-symmetric stable p.d.f. in terms of the Fox's
$H$-function. Thus our study will be focused on the $H$-function
\begin{eqnarray}
H(z)=H^{1, 1}_{2, 2}
\left[ z \Biggl|
\begin{array}{cl} (1,\frac{1}{\alpha}) & (1, \frac{1}{2}) \\
(1, 1) & (1, \frac{1}{2})\end{array} \right]=\frac{1}{2\pi i}\int_{\mathcal{C}}
\frac{\Gamma(1-s)\Gamma(\frac{s}{\alpha})}{\Gamma(\frac{s}{2})\Gamma(1-\frac{s}{2})} z^s ds, \quad z=\frac{|x|}{(K_{\alpha}t)^{\frac{1}{\alpha}}}.
\label{fh3}
\end{eqnarray}
The simple poles of
$\Gamma(\frac{s}{\alpha})$ and  $\Gamma(1-s)$ are given by the disjoint sets of points
\begin{displaymath}
P(s)= \{s_{\nu}=-\alpha \nu, \quad \nu=0,1,\cdots\}
\end{displaymath}
\begin{displaymath}
Q(s)= \{s_{\nu}=1+\nu, \quad \nu=0,1,\cdots\}
\end{displaymath}
We distinguish the following two cases:
\begin{description}
\item[($i$)] Asymptotic expansion of $H(z)$ near the point $z=\infty$. Applying the residue theorem clockwise we find
\begin{eqnarray}
H(z)=\sum_{m=1}^{\infty} Res\{\chi(s)z^s;s_{m}\in P(s)\}+\frac{1}{2\pi i}\int_{\mathcal{C}_1}\chi(s)z^sds-\frac{1}{2\pi i}\int_{\mathcal{C}_2}\chi(s)z^sds
\label{fh31}
\end{eqnarray}
where the contour $\mathcal{C}$ has been replaced by the rectilinear
contours $\mathcal{C}_1, \mathcal{C}_2$ running from $\sigma \rightarrow
\sigma+ik \rightarrow -\infty+ik$ and $\sigma \rightarrow \sigma-ik
\rightarrow -\infty-ik$ respectively. It can be proved that the error terms
on the right-hand side of (\ref{fh31}) vanish. In this case we obtain the
algebraic asymptotic expansion
\begin{eqnarray}
H^{1, 1}_{2, 2}
\left[ z \Biggl|
\begin{array}{cl} (1,\frac{1}{\alpha}) & (1, \frac{1}{2}) \\
(1, 1) & (1, \frac{1}{2})\end{array} \right]=\frac{\alpha}{\pi}\sum_{m=1}^{\infty}
(-1)^{m+1}\frac{\Gamma(1+m\alpha)}{\Gamma(m+1)}\sin(\frac{\pi}{2}m\alpha)z^{-m\alpha}.
\label{fh4}
\end{eqnarray}
It is worth noting that all the points of $P(s)$ contribute in the large z
limit.
\noindent For the series (\ref{fh4}), applying the ratio test, one has
\begin{eqnarray}
\rho_1 =\lim_{m\rightarrow \infty}\left|\frac{\Gamma((m+1)\alpha)}{m\Gamma(m\alpha)}
\frac{\sin(\frac{\pi}{2}(m+1)\alpha)}{\sin(\frac{\pi}{2}m\alpha)}\right|\leq
\lim_{m\rightarrow \infty} \left|\frac{\Gamma((m+1)\alpha)}{m\Gamma(m\alpha)} \right|
= \left\{ \begin{array}{cl} 0 &  \mbox{if} \quad 0<\alpha<1 \\
1 & \mbox{if} \quad \alpha=1 \\
\infty & \mbox{if} \quad 1<\alpha\leq 2. \end{array}\right.
\label{fh51}
\end{eqnarray}
\noindent Thus the series converges absolutely for every value of $z\neq 0$ in the interval
\begin{eqnarray}
(-R_1,R_1)=(-\infty, \infty) \quad \mbox{if} \quad 0<\alpha<1.
\label{fh52}
\end{eqnarray}
\item[($ii$)] Asymptotic expansion of $H(z)$ near the point $z=0$. The function $H(z)$ is analytic in the $s$-plane for $\alpha \in (1,2]$ since then $M=1-\frac{1}{\alpha}>0, \,\,\forall z\neq 0$. Also for $\alpha=1$, $M=0$, and $H$ is analytic for $0<|z|<1$. In this case we find
\begin{eqnarray}
H^{1, 1}_{2, 2}
\left[ z \Biggl|
\begin{array}{cl} (1,\frac{1}{\alpha}) & (1, \frac{1}{2}) \\
(1, 1) & (1, \frac{1}{2})\end{array} \right]&=&-\sum_{m=1}^{\infty} Res\{\chi(s)z^s;s_{m}\in Q(s)\}
= \frac{1}{\pi}\sum_{m=0}^{\infty}
(-1)^{m}\frac{\Gamma(\frac{2m+1}{\alpha})}{\Gamma(2m+1)}z^{2m+1} \nonumber \\
&=& \frac{\alpha}{\pi}\sum_{m=1}^{\infty}(-1)^{m-1}\frac{\Gamma(1+\frac{m}{\alpha})}{\Gamma(1+m)}
\sin(\frac{\pi}{2}m) z^m.
\label{fh5}
\end{eqnarray}
\noindent In this case the even numbers of $Q(s)$ give a vanishing result and
\begin{eqnarray}
\rho_2 =\lim_{m\rightarrow \infty}\left|\frac{\Gamma(\frac{2m+3}{\alpha})}{\Gamma(2m+3)}
\right|
= \left\{ \begin{array}{cl} \infty &  \mbox{if} \quad 0<\alpha<1 \\
1 & \mbox{if} \quad \alpha=1 \\
0 & \mbox{if} \quad 1<\alpha\leq 2. \end{array}\right.
\label{fh53}
\end{eqnarray}
\noindent The series converges absolutely for every value of $z$ in the intervals
\begin{eqnarray}
(-R_2,R_2)=\left\{ \begin{array}{cl}(-1, 1) & \mbox{if} \quad \alpha=1 \\
(-\infty, \infty) & \mbox{if} \quad 1<\alpha\leq 2. \end{array}\right.
\label{fh54}
\end{eqnarray}
\end{description}
We assume now that $\alpha$ is a positive rational number thus it can be
written as: $\alpha=\frac{p}{q}, \,\, p,q\in {\mathbb Z}^+$ (the symbols $p$, $q$
should not be confused with those used in the definition of the
$H$-function). Relation (\ref{fh4}), using the substitution $m=nq+l, \,\,
l=0,\cdots,q-1$, can be casted into the form
\begin{eqnarray}
H(z)=-\frac{\alpha}{\pi}\sum_{l=0}^{q-1}e^{i\pi l}z^{-l\frac{p}{q}}
\left(\sum_{n=0}^{\infty} \frac{\Gamma(np+\frac{lp}{q}+1)}{\Gamma(nq+l+1)}
\sin(\frac{\pi}{2}(n+\frac{l}{q})p)
e^{i\pi nq}z^{-np}\right).
\label{fh6}
\end{eqnarray}
\noindent Using the multiplication theorem of Gauss
\footnote{The multiplication theorem of Gauss states that
$\Gamma(mz)(2\pi)^{(m-1)/2}=m^{mz-1/2}\Gamma(z)\Gamma(z+\frac{1}{m})\cdots
\Gamma(z+\frac{m-1}{m}), \,\, \forall \,m\in\mathcal{N}$.}
 one can write the ratio of the gamma functions as
\begin{eqnarray}
\frac{\Gamma(np+\frac{lp}{q}+1)}{\Gamma(nq+l+1)}=(2\pi)^{\frac{(q-p)}{2}}\sqrt{\frac{p}{q}}
\left(\frac{p^p}{q^q}\right)^{n+\frac{l}{q}}\frac{\prod_{k=0}^{p-1}
\Gamma(\frac{l}{q}+\frac{k+1}{p})}{\prod_{s=0}^{q-1}\Gamma(\frac{l+s+1}{q})}
\frac{\prod_{k=0}^{p-1}(a_k)_n}{\prod_{s=0}^{q-1}(b_s)_n}
\label{fh7}
\end{eqnarray}
\noindent where $(a)_n$ is the Pochhammer's symbol $(a)_n=\frac{\Gamma(a+n)}{\Gamma(a)}$ and
\begin{eqnarray}
a_k=\frac{k+1}{p}+\frac{l}{q}, \quad b_s=\frac{l+s+1}{q}.
\label{fh8}
\end{eqnarray}
\noindent Combining (\ref{fh6}) with (\ref{fh7}) and assuming that $p$ is odd,
we obtain
\begin{eqnarray}
H(z) &=& -\frac{\alpha^{\frac{3}{2}}}{\pi}(2\pi)^{\frac{(q-p)}{2}}
\sum_{l=0}^{q-1}\left(\frac{p^p e^{i\pi q}}{q^q z^p}\right)^{\frac{l}{q}}
\frac{\prod_{k=0}^{p-1}\Gamma(a_k)}{\prod_{s=0}^{q-1}\Gamma(b_s)}
\left( \sin(\frac{\pi lp}{2q})\sum_{m=0}^{\infty}
\frac{\prod_{k=0}^{p-1}(a_k)_{2m}}{\prod_{s=0}^{q-1}(b_s)_{2m}}
 \Biggl(\frac{p^p e^{i\pi (q+\frac{p}{2})}}
{q^q z^p}\right)^{2m} \nonumber \\
&+& e^{-i\frac{\pi}{2}p} \cos(\frac{\pi lp}{2q})\sin({\frac{\pi}{2}p})
\sum_{m=0}^{\infty}
\frac{\prod_{k=0}^{p-1}(a_k)_{2m+1}}{\prod_{s=0}^{q-1}(b_s)_{2m+1}}
 \left(\frac{p^p e^{i\pi (q+\frac{p}{2})}}
{q^q z^p}\right)^{2m+1}\Biggr) \nonumber \\
&=& -\frac{\alpha^{\frac{3}{2}}}{\pi}(2\pi)^{\frac{(q-p)}{2}}
\sum_{l=0}^{q-1}\left(\frac{p^p e^{i\pi q}}{q^q z^p}\right)^{\frac{l}{q}}
\frac{\prod_{k=0}^{p-1}\Gamma(a_k)}{\prod_{s=0}^{q-1}\Gamma(b_s)}
\Biggl( \sin(\frac{\pi lp}{2q})+
e^{-i\frac{\pi}{2}p} \cos(\frac{\pi lp}{2q})\sin({\frac{\pi}{2}p})
\left(\frac{p^p e^{i\pi (q+\frac{p}{2})}}
{q^q z^p}\right)\Biggr)\nonumber \\
&\times&
{}_{p+1}F_{q}\left(1,a_0,\cdots,a_{p-1};b_0,\cdots,b_{q-1};\left(\frac{p^p e^{i\pi (q+\frac{p}{2})}}
{q^q z^p}\right)^{2}\right).
\label{fh9}
\end{eqnarray}
\noindent The same result holds for $p$ even and the only difference is the swapping of the two terms
in the parentheses.
Of course the hypergeometric series converges absolutely when $0<\alpha \leq 1-\frac{1}{q}$ which implies that $\alpha \in (0,1)$.
\par For the case $1< \alpha \leq 2$, following similar steps, we can write the ratio of gamma functions as
\begin{eqnarray}
\frac{\Gamma(\frac{q(2m+1)}{p})}{\Gamma(2m+1)}=\frac{(2\pi)^{p-q}}{2^{(2l+1)(1-\frac{q}{p})}}
\left(\frac{p}{q}\right)^{\frac{1}{2}}\left(\frac{q^{\frac{q}{p}}}{p}\right)^{2l+1}
\left(\frac{(2q)^{2qn}}{(2p)^{2pn}}\right)\frac{\prod_{k=1}^{2q-1}
\Gamma(a_k)}{\prod_{s=1}^{2p-1}\Gamma(b_s)} \frac{\prod_{k=1}^{2q-1}(a_k)_n}{\prod_{s=1}^{2p-1}(b_s)_n}
\label{fh10}
\end{eqnarray}
where $a_k=\frac{1}{2p}(2l+1)+\frac{k}{2q}$ and
$b_s=\frac{1}{2p}(2l+1)+\frac{s}{2p}$. The relation (\ref{fh5}) then becomes
\begin{eqnarray}
H(z)&=& z(2\pi)^{p-q-1}2^{\frac{q}{p}}\frac{q^{\frac{q}{p}-\frac{1}{2}}}{p^{\frac{1}{2}}}
\sum_{l=0}^{p-1}(-1)^l \left(\frac{q^q z^p}{p^p 2^{p-q}}\right)^{\frac{2l}{p}}
\frac{\prod_{k=1}^{2q-1}\Gamma(a_k)}{\prod_{s=1}^{2p-1}
\Gamma(b_s)} \nonumber \\
&\times& {}_{2q}F_{2p-1}\Biggl(1,a_1,\cdots,
a_{2q-1};b_1,\cdots,b_{2p-1}; \frac{(2q)^{2q}}{(2p)^{2p}}e^{i\pi p}z^{2p}\Biggr).
\label{fh11}
\end{eqnarray}
The hypergeometric series ${}_{2q}F_{2p-1}$ converges absolutely if $2q\leq 2p-1$ or
equivalently when
\begin{eqnarray}
\alpha \geq 1+\frac{1}{2q}, \quad q\in {\mathbb Z}^+
\label{fh12}
\end{eqnarray}
\noindent which implies that $\alpha \in (1,2]$. It also converges when
$2q=(2p-1)+1\Rightarrow \alpha=1$ provided that $|z|<1$.
\section{Fox's H-function for the law \boldmath \( S_{\alpha }( \beta,c=K_{\alpha},\tau)\), \boldmath \( \alpha\in (0,2]\) and \boldmath \(\alpha\neq 1 \)}
\label{sec:3}

We consider the generalized one-dimensional anomalous and anisotropic
diffusion problem with the following initial and boundary conditions
\begin{eqnarray}
\frac{\partial f(x,t)}{\partial t}=-\frac{1}{\cos
\left(\frac{\alpha \pi}{2}\right)}\left(K^-\, {}_{-\infty}\mathcal{D}^{\alpha}_x
+K^+\,{}_{x}\mathcal{D}^{\alpha}_{\infty} \right)
f(x,t)-\tau \frac{\partial f(x,t)}{\partial x}
&\equiv& \hat{\mathcal{A}}(\alpha,\beta,\tau)f(x,t),
\nonumber \\ x\in {\mathbb R}, \, t\in(0,\infty),\quad
\lim_{t\downarrow 0}f(x,t)=\delta(x),
&\quad& \lim_{|x|\rightarrow \infty}f(x,t)=0
\label{3a1}
\end{eqnarray}
where $K^{\pm}$ are diffusion constants satisfying $K^{\pm}\geq 0,
K^++K^->0$ and the dispersion term is proportional to the constant $\tau$
having dimensions $[L]/[T]$ \footnote{One might wonder if it is legitimate
to add a Laplacian term to the operator  $\hat{\mathcal{A}}(\alpha,\beta)$.
This suggestion is prohibited by the fact that we consider only
$\alpha$-stable laws.}. Also by definition \cite{Ref23,Ref24}
\begin{eqnarray}
\left({}_{-\infty}\mathcal{D}^{\alpha}_x+{}_{x}\mathcal{D}^{\alpha}_{\infty}\right)f(x,t) \stackrel{{\rm def}}{=} \frac{1}{\Gamma(m-\alpha)}
\left(\frac{\partial}{\partial x}\right)^m
\left( \int_{-\infty}^{x}\frac{f(y,t)}{(x-y)^{\alpha -m +1}}dy +
(-1)^m \int_{x}^{\infty}\frac{f(y,t)}{(y-x)^{\alpha -m +1}}dy \right)
\label{3a2}
\end{eqnarray}
are the nonlocal fractional left-handed (right-handed) Weyl derivatives. In
(\ref{3a2}) $m=[\alpha]+1$ with $[\alpha]$ representing the integral part of
$\alpha$. Note that when $\alpha$ is an even integer then the two
derivatives are localized and equal while for odd integer values of $\alpha$
both derivatives appear opposite in signs.
\par The Fourier transform of $\hat{\mathcal{A}}(\alpha,\beta)f(x,t)$ for fixed $t\in(0,\infty)$ and $f\in \mathcal{S}({\mathbb R})$\footnote{$\mathcal{S}({\mathbb R})$ is the set of all infinitely differentiable and rapidly decreasing functions on ${\mathbb R}$, namely $sup_{x\in {\mathbb R}}|x^n (D^m f)(x)|<\infty, \forall m,n=0,1,\cdots$. This space is usually called the Schwartz space.} is given by (see proof at Appendix A)
\begin{eqnarray}
\mathcal{F}\left[\hat{\mathcal{A}}(\alpha,\beta)f\right](q,t)=\left[-K_{\alpha}|q|^{\alpha}\left(1-i\beta \textrm{sign} (q) \tan\left(\frac{\alpha \pi}{2}\right)\right)+i\tau q\right]\hat{f}(q,t)
\label{3a3}
\end{eqnarray}
where
\begin{eqnarray}
K_{\alpha}=K^++K^->0, \quad \beta=\frac{K^+-K^-}{K^-+K^+}\in[-1,1], \quad \tau\in {\mathbb R}.
\label{3a3a1}
\end{eqnarray}
The operator $\hat{\mathcal{A}}(\alpha,\beta)$ could be casted into the equivalent form
\begin{eqnarray}
\hat{\mathcal{A}}(\alpha,\beta)=-\frac{K_{\alpha}}{2\cos \left(\frac{\alpha \pi}{2}\right)}\left((1-\beta){}_{-\infty}\mathcal{D}^{\alpha}_x +(1+\beta)\,{}_{x}\mathcal{D}^{\alpha}_{\infty} \right)-\tau \frac{\partial }{\partial x}.
\label{3a3a2}
\end{eqnarray}
The $\hat{f}(q,t)$ satisfies the Fourier transformed initial value problem
\begin{eqnarray}
\frac{\partial \hat{f}(q,t)}{\partial t}&=&\left[-K_{\alpha}|q|^{\alpha}\left(1-i\beta \textrm{sign} (q) \tan\left(\frac{\alpha \pi}{2}\right)\right)+i\tau q\right]\hat{f}(q,t) \nonumber \\
\hat{f}(q,0)&=& 1.
\label{3a31}
\end{eqnarray}
with solution
\begin{eqnarray}
\hat{f}(q,t)=e^{-K_{\alpha}|q|^{\alpha}t\left(1-i\beta \textrm{sign} (q) \tan\left(\frac{\alpha \pi}{2}\right)\right)+i\tau q t}.
\label{3a32}
\end{eqnarray}
The propagator is thus given by
\begin{eqnarray}
f(x,t)=\mathcal{F}^{-1}[\hat{f}](x,t)=\frac{1}{2\pi}\int_{-\infty}^{\infty}e^{-iq(x-\tau t)} e^{-K_{\alpha}t|q|^{\alpha}\left(1-i\beta \textnormal{sign}(q) \tan\left(\frac{\alpha \pi}{2}\right)  \right)}dq.
\label{3a4}
\end{eqnarray}
We will first study the small $x-\tau t$ expansion. We expand $\cos(qx), \sin(qx)$ in finite Taylor series
\begin{eqnarray}
\cos(qx)&=& \sum_{n=0}^{m}\frac{(-q^2 x^2)^n}{(2n)!}+\frac{g(qx)(qx)^{2m+1}}{(2m+1)!}, \nonumber \\
\sin(qx)&=& \sum_{n=0}^{m}\frac{(-1)^n(q x)^{2n+1}}{(2n+1)!}+\frac{h(qx)(qx)^{2m+2}}{(2m+2)!}
\label{3a5}
\end{eqnarray}
where, by the generalized mean value theorem, the functions $g(qx), h(qx)$
are bounded by the extreme values of the $(2m+1)$th ($(2m+2)$th) derivative
of the cosine (sine), thus $|g(qx)|,|h(qx)|<1$. By substituting these series
into (\ref{3a4}), integrating term-by-term and taking the limit
$m\rightarrow
\infty$ we obtain the complete asymptotic expansion \footnote{The derivation
of this expression is based on the integral formula
\begin{displaymath}
\int_0^{\infty}x^{\mu -1}e^{-\nu x}\sin(ax)dx=\frac{\Gamma(\mu)}{(\nu^2+a^2)^{\frac{\mu}{2}}}\sin\left(\mu \arctan\frac{a}{\nu}\right), \quad {\rm Re}\mu>-1,\,\, {\rm Re}\nu>|{\rm Im}a|
\end{displaymath}
and a similar result for the cosine, provided that ${\rm Re}\mu>0,\,\, {\rm
Re}\nu>|{\rm Im}a|$.}
\begin{eqnarray}
f(x,t;\alpha,\beta, \tau)=\frac{1}{\pi z \left(K_{\alpha}t\right)^{\frac{1}{\alpha}}\left(1+\gamma^2\right)^{\frac{1}{2\alpha}}} \sum_{n=1}^{\infty} (-1)^{n-1}\frac{\Gamma\left(1+\frac{n}{\alpha}\right)}{\Gamma(n+1)}\sin\left(\frac{n\pi}{2}\delta \right) z^n
\label{3a6}
\end{eqnarray}
where
\begin{eqnarray}
\gamma=\beta \tan\left(\frac{\alpha \pi}{2}\right), \quad \delta=1+\frac{2}{\alpha \pi}\arctan \gamma, \quad \textrm{and} \quad z=\frac{|x-\tau t|}{\left(K_{\alpha}t\right)^{\frac{1}{\alpha}}\left(1+\gamma^2\right)^{\frac{1}{2\alpha}}}.
\label{3a7}
\end{eqnarray}
It is evident from (\ref{3a7}) that $\delta\in[0,2]$ and can acquire the
three integer values $\{0,1,2\}$ provided that $\beta \in \{-1,0,1\}$. The
p.d.f. (\ref{3a6}) can be reproduced by the following H-function
\begin{eqnarray}
H(z)= H^{1, 1}_{2, 2}
\left[ z \Biggl|
\begin{array}{cl} (1,\frac{1}{\alpha}) & (1, \frac{\delta}{2}) \\
(1, 1) & (1, \frac{\delta}{2})\end{array} \right]=\frac{1}{2\pi i}\int_{\mathcal{C}}
\frac{\Gamma(1-s)\Gamma(\frac{s}{\alpha})}{\Gamma(\frac{\delta s}{2})\Gamma(1-\frac{\delta s}{2})} z^s ds,
 \quad z=\frac{|x-\tau t|}{(K_{\alpha}t)^{\frac{1}{\alpha}}\left(1+\gamma^2\right)^{\frac{1}{2\alpha}}}
\label{3a8}
\end{eqnarray}
with asymptotic expansion near the point $z=0$ given by
\begin{eqnarray}
H(z)=\frac{\alpha}{\pi} \sum_{n=1}^{\infty} (-1)^{n-1}\frac{\Gamma\left(1+\frac{n}{\alpha}\right)}{\Gamma(n+1)}\sin\left(\frac{n\pi}{2}\delta \right) z^n, \quad 1<\alpha \leq 2.
\label{3a9}
\end{eqnarray}
In contrast to (\ref{fh5}) all the points of $Q(s)$ now contribute. The
asymptotic expansion of $H(z)$ near the point $z=\infty$ is
\begin{eqnarray}
H(z)=\frac{\alpha}{\pi}\sum_{n=0}^{\infty}(-1)^{n+1}\frac{\Gamma(1+n\alpha)}{\Gamma(1+n)}\sin\left(\frac{\pi n \alpha}{2}\delta \right)|z|^{-n\alpha}, \quad 0<\alpha <1.
\label{3a10}
\end{eqnarray}
For the symmetric $\beta=0$ case we recover expressions (\ref{fh4}) and
(\ref{fh5}). When $\alpha$ is rational we can also express (\ref{3a9}) and
(\ref{3a10}) in terms of hypergeometric functions as follows
\begin{eqnarray}
H(z)&=& -\sqrt{\frac{p}{q}}(2\pi)^{\frac{p-q}{2}-1}\sum_{l=0}^{p-1}\left(\frac{q^{\frac{q}{p}}e^{i\pi}z}{p}\right)^l \frac{\prod_{k=1}^{q-1}\Gamma(a_k)}{\prod_{s=1}^{p-1}\Gamma(b_s)} \nonumber \\
&\times& \Biggl[e^{i\frac{\pi}{2}(l\delta -1)}{}_{q}F_{p-1}\left(1,a_1,\cdots,a_{q-1};b_1,\cdots,b_{p-1}; \frac{e^{i\pi p(1+\frac{\delta}{2})}q^q z^p}{p^p}\right) \nonumber \\
&-& e^{-i\frac{\pi}{2}(l\delta +1)}{}_{q}F_{p-1}\left(1,a_1,\cdots,a_{q-1};b_1,\cdots,b_{p-1}; \frac{e^{i\pi p(1-\frac{\delta}{2})}q^q z^p}{p^p}\right)\Biggr], \quad 1<\alpha\leq 2
\label{3a11}
\end{eqnarray}
with $a_k=\frac{l}{p}+\frac{k}{q}$ and $b_s=\frac{l}{p}+\frac{s}{p}$. Similarly for the large $z$ expansion we obtain
\begin{eqnarray}
H(z)&=& -\left(\frac{p}{q}\right)^{\frac{3}{2}}(2\pi)^{\frac{q-p}{2}-1}\sum_{l=0}^{q-1}\left(\frac{p^p e^{i\pi q}}{q^q |z|^p}\right)^{\frac{l}{q}} \frac{\prod_{k=0}^{p-1}\Gamma(a_k)}{\prod_{s=0}^{q-1}\Gamma(b_s)} \nonumber \\
&\times& \Biggl[e^{i\frac{\pi}{2}(\frac{lp}{q}\delta -1)}{}_{p+1}F_{q}\left(1,a_0,\cdots,a_{p-1};b_0,\cdots,b_{q-1}; \frac{e^{i\pi(q+\frac{p\delta}{2})}p^p}{q^q |z|^p}\right) \nonumber \\
&-& e^{-i\frac{\pi}{2}(\frac{lp}{q}\delta +1)}{}_{p+1}F_{q}\left(1,a_0,\cdots,a_{p-1};b_0,\cdots,b_{q-1}; \frac{e^{i\pi(q-\frac{p\delta}{2})}p^p}{q^q |z|^p}\right)\Biggr], \quad 0<\alpha <1
\label{3a12}
\end{eqnarray}
where $a_k, b_s$ are given by (\ref{fh8}).
\section{The law \boldmath \( \bf S_{\alpha =1}(\beta,c=K,\tau)\)}
\label{sec:4}

The operator we consider in this case is
\begin{eqnarray}
\mathcal{A}(\alpha=1,\beta)f(x,t)=-\frac{d}{dx}\left(K_1(s+\beta h)*f(t))(x)\right)
\label{41}
\end{eqnarray}
where
\begin{eqnarray}
s(x)=\frac{1}{2\pi^2 x}, \quad h(x)=\frac{1}{2\pi^2}\left(\frac{1}{|x|}+2C_{\gamma}\delta(x)\right).
\label{42}
\end{eqnarray}
The first convolution is the Hilbert transform of the function $f$ defined by
\begin{eqnarray}
Hf(x,t)=(s*f(t))(x)=\frac{1}{2\pi^2}\int_{-\infty}^{\infty}f(y,t)s(x-y)dt=\frac{1}{2\pi^2}P.V.\left(\int_{-\infty}^{\infty}\frac{f(y,t)}{x-y}dy\right)
\label{43}
\end{eqnarray}
with $P.V.$ representing the Cauchy principal value of the integral. The
Fourier transform \footnote{We have absorbed the coefficients of the Fourier
transform into the exponential thus defining
\begin{displaymath}\mathcal{F}[f]=\hat{f}(p)=\int_{-\infty}^{\infty}e^{2i\pi
p x}f(x)dx, \quad
\mathcal{F}^{-1}[\hat{f}]=f(x)=\int_{-\infty}^{\infty}e^{-2i\pi p
x}\hat{f}(p)dp. \end{displaymath}} of (\ref{43}) is
\begin{eqnarray}
\mathcal{F}[Hf](x)=\mathcal{F}[s](x)\mathcal{F}[f](x)=\frac{i}{2\pi}sign(q)\hat{f}(q,t).
\label{44}
\end{eqnarray}
The second convolution term in (\ref{41}) has Fourier transform (see
Appendix B for the proof)
\begin{eqnarray}
\mathcal{F}[h*f](x)=\mathcal{F}[h](x)\mathcal{F}[f](x)=-\frac{1}{\pi^2}ln(|q|){f}(q,t).
\label{45}
\end{eqnarray}
Hence, the initial value problem is equivalent to the Fourier transformed
\begin{eqnarray}
\frac{\partial \hat{f}(q,t)}{\partial t}&=&\left[-K|q|\left(1+i\beta \frac{2}{\pi}\textrm{sign} (q) \ln(|q|) \right)+2i\pi q \tau\right]\hat{f}(q,t) \nonumber \\
\hat{f}(q,0)&=& 1
\label{46}
\end{eqnarray}
with solution
\begin{eqnarray}
\label{47}
\hat{f}(q,t)=e^{-K|q|t\left(1+i\beta \frac{2}{\pi}\textrm{sign} (q) \ln(|q|) \right)+2i\pi q \tau t}.
\end{eqnarray}
The propagator is then given by the absolutely convergent integral
\begin{eqnarray}
f(x,t)=\int_{-\infty}^{\infty}e^{-2i\pi q(x-\tau t)}e^{-K|q|t\left(1+i\beta \frac{2}{\pi}\textrm{sign} (q) \ln(|q|) \right)}dq.
\label{48}
\end{eqnarray}
If we set $\beta=0$ in (\ref{48}) we recover the shifted Cauchy p.f.d.
\begin{eqnarray}
f(x,t)=\frac{2K_1t}{(K_1 t)^2+4\pi^2 (x-\tau t)^2}.
\label{48a}
\end{eqnarray}
\section{An alternative way of classifying stable laws}
\label{sec:5}

The symbol of the operator (\ref{3a3a2}) is given by
\begin{eqnarray}
\eta(q)=-K_{\alpha}|q|^{\alpha}\left(1-i\textrm{sign} (q) \gamma \right)+i\tau q.
\label{51}
\end{eqnarray}
From (\ref{51}) the real part ${\rm Re}(\eta(q))\leq 0, \, \forall q \in {\mathbb R}$,
thus we define $h_{\lambda}: \, {\mathbb R}\rightarrow {\mathbb C}, \, \lambda>0$ by
\begin{eqnarray}
h_{\lambda}(q)=\mathcal{L}[e^{t\eta(q)}](\lambda)=\int_0^{\infty}e^{-t \lambda}e^{t \eta(q)} dt=\frac{1}{\lambda-\eta(q)}
\label{52}
\end{eqnarray}
which is positive definite. The mapping $q\rightarrow h_{\lambda}(q)$ is
continuous and applying Bochner's theorem there exists a finite measure on
$\mathcal{B}({\mathbb R})$ such that
\begin{eqnarray}
h_{\lambda}(q)=\widehat{\mu_{\lambda}}(q)=\frac{1}{2\pi}\int_{-\infty}^{\infty}e^{-iqx}\mu_{\lambda}(dx).
\label{53}
\end{eqnarray}
Hence,
\begin{eqnarray}
\mu_{\lambda}(x)=\int_{-\infty}^{\infty}e^{iqx}h_{\lambda}(q)dq.
\label{54}
\end{eqnarray}
It can be shown that the operator $\hat{\mathcal{A}}(\alpha,\beta,\tau)$ is the infinitesimal generator of a strongly continuous semigroup of operators and its resolvent $\mathcal{R}_{\lambda}(\hat{\mathcal{A}})$ satisfies \cite{Ref13}
\begin{eqnarray}
\mathcal{R}_{\lambda}(\hat{\mathcal{A}})\psi =\mu_{\lambda}* \psi,
\label{55}
\end{eqnarray}
or equivalently
\begin{eqnarray}
\left(I-\frac{1}{\lambda}\hat{\mathcal{A}}\right)^{-1}\psi=\lambda \mu_{\lambda}* \psi.
\label{56a}
\end{eqnarray}
By applying on the righthand side of (\ref{56a}) $m$-times the operator
$\left(I-\frac{1}{\lambda}\hat{\mathcal{A}}\right)^{-1}$,  setting
$\lambda=\frac{m}{t}$ and taking the limit $m\rightarrow \infty$ we have
\begin{eqnarray}
\lim_{m\rightarrow \infty}\left(I-\frac{t }{m}\hat{\mathcal{A}}\right)^{-m}\psi &=& e^{-t \hat{\mathcal{A}}} \psi=\lim_{m\rightarrow \infty} \left( \left(\frac{m}{t}\right)^m \underbrace{\mu_{m/t}*\cdots*\mu_{m/t}}_{m-{\rm times}}\right)* \psi \nonumber \\
&=& f*\psi
\label{56b}
\end{eqnarray}
where $f$ is the p.d.f. of the stable law $S_{\alpha}(\beta,
K_{\alpha},\tau)$. Thus, a stable law can be determined by $\mu_{\lambda}$
instead of $f$, using relations (\ref{52}) and (\ref{54}). As an example we
consider the law $S_2(0,K_2,0)$. A simple calculation gives the p.d.f.
\begin{eqnarray}
\mu_{\lambda}(x)=\frac{\pi}{2\sqrt{K_2 \lambda}}e^{-\sqrt{\frac{\lambda}{K_2}}x}, \quad x>0.
\label{56c}
\end{eqnarray}
Although (\ref{54}) seems to be elegant it is unattractive for calculations
since the corresponding convergent integral only exceptionally gives a
closed expression.
\section{Known and unknown results}
\label{sec:6}

Expressions (\ref{fh9}), (\ref{fh11}), (\ref{3a11}) and (\ref{3a12}) can be
used to calculate p.d.f.'s for arbitrary rational values of $\alpha \in
(0,2]$ in terms of generalized hypergeometric functions. In particular, a
sample of stable symmetric p.d.f.'s is given in Table 1 for the subdiffusion
regime with $\alpha \in (0,1)$ and in Table 2 for the superdiffusion regime
with $\alpha \in (1,2]$. Previous known results are reproduced and new ones
are also presented.
\begin{description}
\item[$\textrm{\boldmath{$(\alpha)$}}$] \textbf{The law} $\mathbf{\textrm{\boldmath{$S_{1/2}(0,K_{1/2},0)$}}}$
\par Let us first consider the particular case $(p,q)=(1,2)$ that gives the index value $\alpha=1/2$. From equation (\ref{fh9}) one can easily show that
\begin{eqnarray}
H(z)&=& -\frac{1}{2\sqrt{\pi}}\sum_{l=0}^{1} \left(\frac{e^{i\pi}}{2\sqrt{z}}\right)^l
\left(\sum_{n=0}^{\infty}\frac{\sin(\frac{\pi}{2}(n+\frac{l}{2}))}{\Gamma(n+\frac{(l+1)}{2})}
\frac{1}{(4z)^n}\right)\nonumber \\
&=&-\frac{1}{4\pi z}\left[ \cos(\frac{1}{4z})\,\,
{}_{1}F_2(\frac{1}{4};\frac{1}{2},\frac{5}{4},-\frac{1}{64z^2})+\frac{1}{12z}\sin(\frac{1}{4z})\,\,
{}_{1}F_2(\frac{3}{4};\frac{3}{2},\frac{7}{4},-\frac{1}{64z^2})\right]
\nonumber \\
&+& \frac{\sqrt{2}}{8\sqrt{\pi z}}\left[\frac{1}{4z}\,\,{}_0F_1(\frac{3}{2};-\frac{1}{64z^2})
+{}_0F_1(\frac{1}{2};-\frac{1}{64z^2})\right] \nonumber \\
&=& -\frac{1}{2\sqrt{2\pi |z|}}\left[\cos(\frac{1}{4z}) C(\frac{1}{\sqrt{2\pi z}})+
\sin(\frac{1}{4z}) S(\frac{1}{\sqrt{2\pi z}})\right] \nonumber \\
&+& \frac{1}{4\sqrt{2\pi z}}\left[\cos(\frac{1}{4z})+\sin(\frac{1}{4z}) \right].
\label{ku6}
\end{eqnarray}
\noindent Where $C(x), S(x)$ are the cosine and sine Fresnel integrals given by
\begin{eqnarray}
C(x)&=&\int_{0}^{x}\cos(\frac{\pi}{2}t^2)dt=x \,\,
{}_{1}F_2(\frac{1}{4};\frac{1}{2},\frac{5}{4},-\frac{\pi^2}{16}x^4)\nonumber \\
&=& x\frac{\Gamma(\frac{1}{2})}{\Gamma(\frac{1}{4})^2}\int_{0}^{1}t^{-\frac{3}{4}}(1-t)^{-\frac{3}{4}}\,\, {}_{0}F_1(\frac{5}{4},-\frac{\pi^2}{16}x^4 t) dt \nonumber \\
S(x)&=& \int_{0}^{x}\sin(\frac{\pi}{2}t^2)dt=\frac{\pi}{6}x^3 \,\,
{}_{1}F_2(\frac{3}{4};\frac{3}{2},\frac{7}{4},-\frac{\pi^2}{16}x^4) \nonumber \\
&=& x\frac{\Gamma(\frac{3}{2})}{\Gamma(\frac{3}{4})^2}\int_{0}^{1}t^{-\frac{1}{4}}(1-t)^{-\frac{1}{4}}\,\,{}_{0}F_1(\frac{7}{4},-\frac{\pi^2}{16}x^4 t) dt
\label{ku7}
\end{eqnarray}
\noindent which are odd functions of $x$. The function $f(z)=\frac{1}{\alpha z}H(z)$ is indeed a p.d.f. since it is positive definite $\forall z\in(-\infty,\infty)$, integrable and satisfies
\begin{eqnarray}
\int_{-\infty}^{\infty}f(z)dz=1.
\label{ku8}
\end{eqnarray}
The more general one-sided law $S_{1/2}(1,K_{1/2},\tau)$ is found to correspond to the p.d.f.
\begin{eqnarray}
f(z)=\frac{1}{2 z^{\frac{3}{2}}\sqrt{\pi}} e^{-\frac{1}{4 z}}, \quad z>0.
\label{ku81}
\end{eqnarray}
The case $\beta=-1$ gives a vanishing p.d.f. since $\delta=0$. This result holds independently of the value of $\alpha$.
\item[$\textrm{\boldmath{$(\beta)$}}$] \textbf{The Cauchy law} $\mathbf{\textrm{\boldmath{$S_{1}(0,K_{1},0)$}}}$
\par Setting in (\ref{fh11}), $p=q$, yields
\begin{eqnarray}
H(z)&=&\frac{z}{\pi}{}_{1}F_{0}(1;0;(-z^2)^q)
\sum_{l=0}^{q-1}(-1)^l z^{2l}\nonumber \\
&=& \frac{z}{\pi}\left(\frac{1-(-z^2)^q}{1+z^2}\right){}_{1}F_{0}(1;0;(-z^2)^q).
\label{ku1}
\end{eqnarray}
\noindent When $q=1$ this corresponds to the Cauchy p.d.f.
\begin{eqnarray}
f(z)=\frac{1}{\alpha z}H(z)=\frac{1}{\pi}\frac{1}{(1+z^2)}, \quad z\in(-\infty,\infty).
\label{ku2}
\end{eqnarray}
\noindent with applications , e.g. in spectroscopy. For the general law $S_1(0,K_1,\tau)$ the p.d.f. is given by (\ref{48a}).
\item[$\textrm{\boldmath{$(\gamma)$}}$] \textbf{The Holtsmark law} $\mathbf{\textrm{\boldmath{$S_{3/2}(0,K_{3/2},0)$}}}$
\par In this case we set $(p,q)=(3,2)$ in (\ref{fh11}) and arrive at
\begin{eqnarray}
H(z)&=& \frac{z 2^{\frac{5}{6}}}{\sqrt{3}}\sum_{l=0}^{2}(-1)^l
\left(\frac{2}{27}z^3\right)^{\frac{2l}{3}}
\frac{\prod_{k=1}^{3}\Gamma(\frac{1}{6}(2l+1)+\frac{k}{4})}{\prod_{s=1}^{5}
\Gamma(\frac{1}{6}(2l+1)+\frac{s}{6})}
\nonumber \\
&\times& {}_{4}F_{5}\Biggl(1,\frac{1}{6}(2l+1)+\frac{1}{4},\cdots,\frac{1}{6}(2l+1)+
\frac{3}{4};\frac{1}{6}(2l+1)+\frac{1}{6},\cdots,\frac{1}{6}(2l+1)+\frac{5}{6}\nonumber \\
&;& \frac{4^{4}}{6^{6}}e^{3i\pi}z^{6}\Biggr)\nonumber \\
&=& \frac{z}{\pi}\Biggl[\Gamma(\frac{2}{3})\,\,
{}_2F_3(\frac{5}{12},\frac{11}{12};\frac{1}{3},\frac{1}{2},\frac{5}{6};-\frac{4}{729}
z^6)-\frac{z^2}{2}\,\,
{}_3F_4(1,\frac{3}{4},\frac{5}{4};\frac{5}{6},\frac{2}{3},\frac{4}{3},
\frac{7}{6};-\frac{4}{729}z^6) \nonumber \\
&+& \frac{14\sqrt{3}}{486}\frac{\pi}{\Gamma(\frac{2}{3})}z^4 \,\,
{}_2F_3(\frac{19}{12},\frac{13}{12};\frac{3}{2},\frac{7}{6},\frac{5}{3}
;-\frac{4}{729}z^6)\Biggr].
\label{ku3}
\end{eqnarray}
The corresponding p.d.f. was discovered in physics by the Danish astronomer
Holtsmark in 1919 \cite{Ref26,Ref27}. It was the outcome of his efforts to
study the stationary distribution of the force acting on a star, per unit
mass, due to the gravitational attraction of the neighboring stars.
\item[$\textrm{\boldmath{$(\delta)$}}$] \textbf{The Gaussian law} $\mathbf{\textrm{\boldmath{$S_{2}(1,K_{2},\tau)$}}}$
\par Substituting $p=2q$ in (\ref{fh11}) we get
\begin{eqnarray}
&H(z)&= \frac{z(2\pi)^{q-1}}{\sqrt{q}} \sum_{l=0}^{2q-1}(-1)^l
\left(\frac{z^2}{8q}\right)^l \frac{\prod_{k=1}^{2q-1}
\Gamma(\frac{2(l+k)+1}{4q})}
{\prod_{s=1}^{4q-1}\Gamma(\frac{2l+s+1}{4q})} \nonumber \\
&\times&\!\!\!\!\!
{}_{2q}F_{4q-1}\Biggl(1,a_1,\cdots,a_{2q-1};b_1,\cdots,b_{2p-1};
\frac{(2q)^{2q}}{(4q)^{4q}}e^{2i\pi q}z^{4q}\Biggr)\nonumber \\
&=& \frac{z(2\pi)^{q-1}}{\sqrt{q}} \sum_{l=0}^{2q-1}(-1)^l
\left(\frac{z^2}{8q}\right)^l \frac{1}
{\prod_{s=0}^{2q-1}\Gamma(\frac{l+s+1}{2q})} \nonumber \\
&\times&\!\!\!\!\!
{}_{2q}F_{4q-1}\Biggl(1,a_1,\cdots,a_{2q-1};b_1,\cdots,b_{2p-1};
\frac{1}{(4^3q^2)^q}z^{4q}\Biggr)
\label{ku4}
\end{eqnarray}
\noindent where $a_k=\frac{2l+1}{4q}+\frac{k}{2q}$ and $b_s=\frac{2l+1}{2p}+\frac{s}{2p}$.
\noindent Setting $q=1$ in (\ref{ku4}) we have
\begin{eqnarray}
H(z)&=& \frac{z}{\sqrt{\pi}}\left[{}_0F_1(\frac{1}{2};\frac{z^4}{64})-\frac{z^2}{4}
{}_0F_1(\frac{3}{2};\frac{z^4}{64})\right]
= \frac{z}{\sqrt{\pi}}\left(\cosh(\frac{z^2}{4})-\sinh(\frac{z^2}{4})\right) \nonumber \\
&=& \frac{z}{\sqrt{\pi}}e^{-\frac{z^2}{4}}.
\label{ku5}
\end{eqnarray}
\noindent The associated p.d.f. is
\begin{displaymath}
f(z)=\frac{1}{2\sqrt{\pi}}e^{-\frac{z^2}{4}}
\end{displaymath}
and changing variable to $z=\frac{|x-\tau t|}{\sqrt{K_2 t}}$ we recover the traditional one
\begin{displaymath}
f(x,t)= \frac{1}{\sqrt{4\pi K_2 t}}e^{-\frac{(x-\tau t)^2}{4K_2t}}.
\end{displaymath}
\end{description}
\clearpage
\renewcommand{\baselinestretch}{1}
\begin{table}[!h]
\caption{The Fox's $H$-function for the Farey series $\mathcal{F}_n$ of
order $n=5$ excluding the first $\frac{0}{1}$ and the last $\frac{1}{1}$
member of the series. The functions $csc(z)$ and $I_{\nu}(z)$ are the
cosecant of $z$ and the modified Bessel function of the first kind given by:
$I_{\nu}(z)=e^{-i\pi\frac{\nu}{2}}J_{\nu}(e^{i\frac{\pi}{2}}z)=\frac{1}{\Gamma(\nu+1)}\left(\frac{z}{2}\right)^{\nu}
{}_{0}F_1(\nu+1;\frac{z^2}{4})$.}
\centering
\label{tab:1}
{\small
\begin{tabular}{|c||c|} \hline
$\alpha$  & $H(z)=\alpha |z| f(z)$  \\
\hline\hline
$\frac{1}{5}$     &  \begin{math}
\frac{1}{25}\frac{csc(\frac{\pi}{5})\sin(\frac{\pi}{10})}{z^{\frac{1}{5}}\Gamma(\frac{4}{5})}
{}_{0}F_{7}(\frac{1}{5},\frac{3}{10},\frac{2}{5},\frac{1}{2},\frac{7}{10},\frac{4}{5},
\frac{9}{10};-\frac{1}{25\cdot 10^{8}z^2})\end{math}\\
& \begin{math}
-\frac{1}{25}\frac{csc(\frac{2\pi}{5})\sin(\frac{\pi}{5})}{z^{\frac{2}{5}}\Gamma(\frac{3}{5})}
{}_{0}F_{7}(\frac{3}{10},\frac{2}{5},\frac{1}{2},\frac{3}{5},\frac{4}{5},\frac{9}{10},
\frac{11}{10};-\frac{1}{25\cdot 10^{8}z^2}) \end{math}\\
& \begin{math}
+\frac{1}{50}\frac{\Gamma(\frac{3}{5})\sin(\frac{3\pi}{10})}{z^{\frac{3}{5}}\pi}
{}_{0}F_{7}(\frac{2}{5},\frac{1}{2},\frac{3}{5},\frac{7}{10},\frac{9}{10},
\frac{11}{10}, \frac{6}{5};-\frac{1}{25\cdot 10^{8}z^2})\end{math}\\
& \begin{math}
-\frac{1}{150}\frac{\Gamma(\frac{4}{5})\sin(\frac{2\pi}{5})}{z^{\frac{4}{5}}\pi}
{}_{0}F_{7}(\frac{1}{2},\frac{3}{5},\frac{7}{10},\frac{4}{5},
\frac{11}{10}, \frac{6}{5},\frac{13}{10};-\frac{1}{25\cdot 10^{8}z^2})
\end{math} \\
& \begin{math}
+\frac{1}{600}\frac{1}{z\pi}
{}_{1}F_{8}(1;\frac{3}{5},\frac{7}{10},\frac{4}{5},\frac{9}{10},
\frac{11}{10}, \frac{6}{5},\frac{13}{10},\frac{7}{5};-\frac{1}{25\cdot 10^{8}z^2})\end{math}\\
& \begin{math}
-\frac{1}{15000}\frac{csc(\frac{\pi}{5})\sin(\frac{2\pi}{5})}{z^{\frac{6}{5}}\Gamma(\frac{4}{5})}
{}_{0}F_{7}(\frac{7}{10},\frac{4}{5},\frac{9}{10},\frac{6}{5},
\frac{13}{10}, \frac{7}{5},\frac{3}{2};-\frac{1}{25\cdot 10^{8}z^2})
\end{math} \\
& \begin{math}
+ \frac{1}{45000}\frac{csc(\frac{2\pi}{5})\sin(\frac{3\pi}{10})}{z^{\frac{7}{5}}
\Gamma(\frac{3}{5})}{}_{0}F_{7}(\frac{4}{5},\frac{9}{10},\frac{11}{10},
\frac{13}{10}, \frac{7}{5},\frac{3}{2},\frac{8}{5};-\frac{1}{25\cdot 10^{8}z^2})
\end{math}\\
& \begin{math}
- \frac{1}{210000}\frac{\Gamma(\frac{3}{5})\sin(\frac{\pi}{5})}{z^{\frac{8}{5}}\pi}
{}_{0}F_{7}(\frac{9}{10},\frac{11}{10},\frac{6}{5},
\frac{7}{5},\frac{3}{2},\frac{8}{5},\frac{17}{10};-\frac{1}{25\cdot 10^{8}z^2})
\end{math}\\
& \begin{math}
+ \frac{1}{1260000}\frac{\Gamma(\frac{4}{5})\sin(\frac{\pi}{10})}{z^{\frac{9}{5}}\pi}
{}_{0}F_{7}(\frac{11}{10},\frac{6}{5},\frac{13}{10},
\frac{3}{2},\frac{8}{5},\frac{17}{10},\frac{9}{5};-\frac{1}{25\cdot 10^{8}z^2})
\end{math}\\
\hline
$\frac{1}{4}$ & \begin{math}
\frac{1}{16}\frac{\sqrt{2}\sin(\frac{\pi}{8})}{z^{\frac{1}{4}}\Gamma(\frac{3}{4})}
{}_{0}F_{5}(\frac{1}{4},\frac{3}{8},\frac{1}{2},\frac{3}{4},
\frac{7}{8};-\frac{1}{4194304 z^2})
-\frac{1}{32}\frac{\sqrt{2})}{\sqrt{z}\sqrt{\pi}}
{}_{0}F_{5}(\frac{3}{8},\frac{1}{2},\frac{5}{8},
\frac{7}{8},\frac{9}{8};-\frac{1}{4194304 z^2})
\end{math}\\
& \begin{math}
+\frac{1}{32}\frac{\Gamma(\frac{3}{4})\sin(\frac{3\pi}{8})}{z^{\frac{3}{4}}\pi}
{}_{0}F_{5}(\frac{1}{2},\frac{5}{8},\frac{3}{4},
\frac{9}{8},\frac{5}{4};-\frac{1}{4194304 z^2})
-\frac{1}{96}\frac{1}{z \pi}
{}_{1}F_{6}(1;\frac{5}{8},\frac{3}{4},\frac{7}{8},
\frac{9}{8},\frac{5}{4},\frac{11}{8};-\frac{1}{4194304 z^2})
\end{math} \\
& \begin{math}
+\frac{1}{1536}\frac{\sqrt{2}\sin(\frac{3\pi}{8})}{z^{\frac{5}{4}}\Gamma(\frac{3}{4})}
{}_{0}F_{5}(\frac{3}{4},\frac{7}{8},\frac{5}{4},
\frac{11}{8},\frac{3}{2};-\frac{1}{4194304 z^2})
-\frac{1}{7680}\frac{\sqrt{2}}{z^{\frac{3}{2}}\sqrt{\pi}}
{}_{0}F_{5}(\frac{7}{8},\frac{9}{8},
\frac{11}{8},\frac{3}{2},\frac{13}{8};-\frac{1}{4194304 z^2})\end{math}\\
& \begin{math}
+ \frac{1}{15360}\frac{\Gamma(\frac{3}{4})\sin(\frac{\pi}{8})}{z^{\frac{7}{4}}\pi}
{}_{0}F_{5}(\frac{9}{8},\frac{5}{4},
\frac{3}{2},\frac{13}{8},\frac{7}{4};-\frac{1}{4194304 z^2})\end{math}
\\
\hline
$\frac{1}{3}$ & \begin{math}
\frac{1}{27}\frac{\sqrt{3}}{z^{\frac{1}{3}}\Gamma(\frac{2}{3})}
{}_{0}F_{3}(\frac{1}{3},\frac{1}{2},\frac{5}{6};-\frac{1}{11664 z^2})
-\frac{1}{18}\frac{\sqrt{3}\Gamma(\frac{2}{3})}{z^{\frac{2}{3}}\pi}
{}_{0}F_{3}(\frac{1}{2},\frac{2}{3},\frac{7}{6};-\frac{1}{11664 z^2})\end{math}\\
& \begin{math}
+ \frac{1}{18}\frac{1}{z\pi}
{}_{1}F_{4}(1;\frac{2}{3},\frac{5}{6},\frac{7}{6},\frac{4}{3};-\frac{1}{11664 z^2})
\end{math}\\
& \begin{math}
-\frac{1}{162}\frac{1}{z^{\frac{4}{3}}\Gamma(\frac{2}{3})}
{}_{0}F_{3}(\frac{5}{6},\frac{4}{3},\frac{3}{2};-\frac{1}{11664 z^2})
+\frac{1}{648}\frac{\Gamma(\frac{2}{3})}{z^{\frac{5}{3}}\pi}
{}_{0}F_{3}(\frac{7}{6},\frac{3}{2},\frac{5}{3};-\frac{1}{11664 z^2})
\end{math}\\
\hline
$\frac{2}{5}$ & \begin{math}
\frac{4}{25}\frac{1}{z^{\frac{2}{5}}\Gamma(\frac{3}{5})}csc(\frac{2\pi}{5})\sin(\frac{\pi}{5})
{}_{2}F_{7}(\frac{7}{20},\frac{17}{20};\frac{1}{5},\frac{3}{10},\frac{2}{5},\frac{1}{2},\frac{7}{10},\frac{4}{5},\frac{9}{10};\frac{1}{39062500 z^4})
\end{math} \\
& \begin{math}
-\frac{4}{25}\Gamma(\frac{4}{5})\sin(\frac{2\pi}{5})\frac{1}{z^{\frac{4}{5}}\pi}
{}_{2}F_{7}(\frac{9}{20},\frac{19}{20};\frac{3}{10},\frac{2}{5},\frac{1}{2},\frac{3}{5},\frac{4}{5},\frac{9}{10},\frac{11}{10};\frac{1}{39062500 z^4})
\end{math} \\
& \begin{math}
+\frac{2}{125}\frac{csc(\frac{\pi}{5})\sin(\frac{2\pi}{5})}{z^{\frac{6}{5}}\Gamma(\frac{4}{5})}
{}_{2}F_{7}(\frac{11}{20},\frac{21}{20};\frac{2}{5},\frac{1}{2},\frac{3}{5},\frac{7}{10},\frac{9}{10},\frac{11}{10},\frac{6}{5};\frac{1}{39062500 z^4})
\end{math}\\
& \begin{math}
-\frac{2}{125}\frac{\Gamma(\frac{3}{5})\sin(\frac{\pi}{5})}{z^{\frac{8}{5}}\pi}
{}_{2}F_{7}(\frac{13}{20},\frac{23}{20};\frac{1}{2},\frac{3}{5},\frac{7}{10},\frac{4}{5},\frac{11}{10},\frac{6}{5},\frac{13}{10};\frac{1}{39062500 z^4})
\end{math} \\
& \begin{math}
+\frac{7}{9375}\frac{csc(\frac{2\pi}{5})\sin(\frac{\pi}{5})}{z^{\frac{12}{5}}\Gamma(\frac{3}{5})}
{}_{2}F_{7}(\frac{17}{20},\frac{27}{20};\frac{7}{10},\frac{4}{5},\frac{9}{10},\frac{6}{5},\frac{13}{10},\frac{7}{5},\frac{3}{2};\frac{1}{39062500 z^4})
\end{math}\\
& \begin{math}
-\frac{1}{3125}\frac{\Gamma(\frac{4}{5})\sin(\frac{2\pi}{5})}{z^{\frac{14}{5}}\pi}
{}_{2}F_{7}(\frac{19}{20},\frac{29}{20};\frac{4}{5},\frac{9}{10},\frac{11}{10},\frac{13}{10},\frac{7}{5},\frac{3}{2},\frac{8}{5};\frac{1}{39062500 z^4})
\end{math}\\
& \begin{math}
+\frac{11}{656250}\frac{csc(\frac{\pi}{5})\sin(\frac{2\pi}{5})
}{z^{\frac{16}{5}}\Gamma(\frac{4}{5})}
{}_{2}F_{7}(\frac{21}{20},\frac{31}{20};\frac{9}{10},\frac{11}{10},\frac{6}{5},\frac{7}{5},\frac{3}{2},\frac{8}{5},\frac{17}{10};\frac{1}{39062500 z^4})
\end{math} \\
& \begin{math}
-\frac{13}{1312500}\frac{\Gamma(\frac{3}{5})\sin(\frac{\pi}{5})}{z^{\frac{18}{5}}\pi}
{}_{2}F_{7}(\frac{23}{20},\frac{33}{20};\frac{11}{10},\frac{6}{5},\frac{13}{10},\frac{3}{2},\frac{8}{5},\frac{17}{10},\frac{9}{5};\frac{1}{39062500 z^4})
\end{math} \\ 
\hline
$\frac{1}{2}$ & \begin{math}
-\frac{1}{2\sqrt{2\pi z}}\left[\cos(\frac{1}{4z}) C(\frac{1}{\sqrt{2\pi z}})+
\sin(\frac{1}{4z}) S(\frac{1}{\sqrt{2\pi z}})\right]
+\frac{1}{4\sqrt{2\pi z}}\left[\cos(\frac{1}{4z})+\sin(\frac{1}{4z}) \right].
\end{math}\\
\hline
\end{tabular}
}
\end{table}
\newpage
\begin{table}[!h]
\centering
{\small
\begin{tabular}{|c||c|} \hline
$\frac{3}{5}$ & \begin{math}
\frac{9}{25}\frac{\Gamma(\frac{3}{5})\sin(\frac{3\pi}{10})}{z^{\frac{3}{5}}\pi}
{}_{4}F_{7}(\frac{4}{15},\frac{13}{30},\frac{23}{30},\frac{14}{15};\frac{1}{5},
\frac{3}{10},\frac{2}{5},\frac{1}{2},\frac{7}{10},\frac{4}{5},\frac{9}{10};
-\frac{729}{15625\cdot 10^4 z^6})
\end{math}\\
& \begin{math}
-\frac{9}{125}\frac{csc(\frac{\pi}{5})\sin(\frac{2\pi}{5})}{z^{\frac{6}{5}}\Gamma(\frac{4}{5})}
{}_{4}F_{7}(\frac{11}{30},\frac{8}{15},\frac{13}{15},\frac{31}{30};\frac{3}{10},
\frac{2}{5},\frac{1}{2},\frac{3}{5},\frac{4}{5},\frac{9}{10},\frac{11}{10};
-\frac{729}{15625\cdot 10^4 z^6})
\end{math}\\
& \begin{math}
+\frac{18}{125}\frac{\Gamma(\frac{4}{5})\sin(\frac{\pi}{10})}{z^{\frac{9}{5}}\pi}
{}_{4}F_{7}(\frac{7}{15},\frac{19}{30},\frac{29}{30},\frac{17}{15};\frac{2}{5},
\frac{1}{2},\frac{3}{5},\frac{7}{10},\frac{9}{10},\frac{11}{10},\frac{6}{5};
-\frac{729}{15625\cdot 10^4 z^6})
\end{math}\\
& \begin{math}
+\frac{21}{625}\frac{csc(\frac{2\pi}{5})\sin(\frac{\pi}{5})}{z^{\frac{12}{5}}\Gamma(\frac{3}{5})}
{}_{4}F_{7}(\frac{17}{30},\frac{11}{15},\frac{16}{15},\frac{37}{30};\frac{1}{2},
\frac{3}{5},\frac{7}{10},\frac{4}{5},\frac{11}{10},\frac{6}{5},\frac{13}{10};
-\frac{729}{15625\cdot 10^4 z^6})
\end{math}\\
& \begin{math}
-\frac{3}{100}\frac{1}{z^3 \pi}
{}_{5}F_{8}(\frac{2}{3},\frac{5}{6},1,\frac{7}{6},\frac{4}{3};\frac{3}{5},
\frac{7}{10},\frac{4}{5},\frac{9}{10},\frac{11}{10},\frac{6}{5},\frac{13}{10},\frac{7}{5};
-\frac{729}{15625\cdot 10^4 z^6})
\end{math}\\
& \begin{math}
+\frac{117}{15625}\frac{\Gamma(\frac{3}{5})\sin(\frac{\pi}{5})}{z^{\frac{18}{5}}\pi}
{}_{4}F_{7}(\frac{23}{30},\frac{14}{15},\frac{19}{15},\frac{43}{30};\frac{7}{10},
\frac{4}{5},\frac{9}{10},\frac{6}{5},\frac{13}{10},\frac{7}{5},\frac{3}{2};
-\frac{729}{15625\cdot 10^4 z^6})
\end{math}\\
& \begin{math}
+\frac{66}{78125}\frac{csc(\frac{\pi}{5})\sin(\frac{\pi}{10})}{z^{\frac{21}{5}}\Gamma(\frac{4}{5})}
{}_{4}F_{7}(\frac{13}{15},\frac{31}{30},\frac{41}{30},\frac{23}{15};\frac{4}{5},
\frac{9}{10},\frac{11}{10},\frac{13}{10},\frac{7}{5},\frac{3}{2},\frac{8}{5};
-\frac{729}{15625\cdot 10^4 z^6})
\end{math}\\
& \begin{math}
-\frac{171}{156250}\frac{\Gamma(\frac{4}{5})\sin(\frac{2\pi}{5})}{z^{\frac{24}{5}}\pi}
{}_{4}F_{7}(\frac{29}{30},\frac{17}{15},\frac{22}{15},\frac{49}{30};\frac{9}{10},
\frac{11}{10},\frac{6}{5},\frac{7}{5},\frac{3}{2},\frac{8}{5},\frac{17}{10};
-\frac{729}{15625\cdot 10^4 z^6})
\end{math}\\
& \begin{math}
+\frac{561}{31125000}\frac{csc(\frac{2\pi}{5})\sin(\frac{3\pi}{10})}{z^{\frac{27}{5}}\Gamma(\frac{3}{5})}
{}_{4}F_{7}(\frac{16}{15},\frac{37}{30},\frac{47}{30},\frac{26}{15};\frac{11}{10},
\frac{6}{5},\frac{13}{10},\frac{3}{2},\frac{8}{5},\frac{17}{10},\frac{9}{5};
-\frac{729}{15625\cdot 10^4 z^6})
\end{math}\\
\hline
$\frac{2}{3}$ & \begin{math}
\frac{4}{9}\left(\cosh(\frac{2}{27z^2})I_{\frac{1}{3}}(\frac{2}{27z^2})+\frac{1}{9z^2}\sinh(\frac{2}{27z^2})I_{\frac{1}{3}}(\frac{2}{27z^2})+\frac{1}{9z^2}\cosh(\frac{2}{27z^2})I_{\frac{4}{3}}(\frac{2}{27z^2}) \right)
\end{math}\\
& \begin{math}
-\frac{4}{27}\left(6\cosh(\frac{2}{27z^2}I_{\frac{2}{3}}(\frac{2}{27z^2}))+\frac{1}{3z^2}\sinh(\frac{2}{27z^2})I_{\frac{2}{3}}(\frac{2}{27z^2})+\frac{1}{3z^2}\cosh(\frac{2}{27z^2})I_{\frac{5}{3}}(\frac{2}{27z^2}) \right)
\end{math}\\
& \begin{math}
+\frac{4}{81z^2}\left(\cosh(\frac{2}{27z^2})I_{\frac{1}{3}}(\frac{2}{27z^2})+9z^2 \sinh(\frac{2}{27z^2})I_{\frac{1}{3}}(\frac{2}{27z^2})+\sinh(\frac{2}{27z^2})I_{\frac{4}{3}}(\frac{2}{27z^2}) \right)
\end{math}\\
& \begin{math}
-\frac{4}{81z^2}\left(\cosh(\frac{2}{27z^2})I_{\frac{2}{3}}(\frac{2}{27z^2})+18z^2\sinh(\frac{2}{27z^2})I_{\frac{2}{3}}(\frac{2}{27z^2})+\sinh(\frac{2}{27z^2})I_{\frac{5}{3}}(\frac{2}{27z^2})\right)
\end{math}\\
\hline
$\frac{3}{4}$ & \begin{math}
\frac{9}{16}\frac{\Gamma(\frac{3}{4})\sin(\frac{3\pi}{8})}{z^{\frac{3}{4}}\pi}
{}_{4}F_{5}(\frac{7}{24},\frac{11}{24},\frac{19}{24},\frac{23}{24};\frac{1}{4},\frac{3}{8},\frac{1}{2},\frac{3}{4},\frac{7}{8};-\frac{729}{262144z^6})
\end{math}\\
& \begin{math}
-\frac{9}{64}\frac{\sqrt{2}}{z^{\frac{3}{2}}\sqrt{\pi}}
{}_{4}F_{5}(\frac{5}{12},\frac{7}{12},\frac{11}{12},\frac{13}{12};\frac{3}{8},\frac{1}{2},\frac{5}{8},\frac{7}{8},\frac{9}{8};-\frac{729}{262144z^6})
\end{math}\\
& \begin{math}
-\frac{45}{512}\frac{\sqrt{2}\sin(\frac{\pi}{8})}{z^{\frac{9}{4}}\Gamma(\frac{3}{4})}
{}_{4}F_{5}(\frac{13}{24},\frac{17}{24},\frac{25}{24},\frac{29}{24};\frac{1}{2},\frac{5}{8},\frac{3}{4},\frac{9}{8},\frac{5}{4};-\frac{729}{262144z^6})
\end{math}\\
& \begin{math}
+\frac{3}{16}\frac{1}{z^3 \pi}
{}_{5}F_{6}(\frac{2}{3},\frac{5}{6},1,\frac{7}{6},\frac{4}{3};\frac{5}{8},\frac{3}{4},\frac{7}{8},\frac{9}{8},\frac{5}{4},\frac{11}{8};-\frac{729}{262144z^6})
\end{math}\\
& \begin{math}
-\frac{693}{8192}\frac{\Gamma(\frac{3}{4})\sin(\frac{\pi}{8})}{z^{\frac{15}{4}}\pi}
{}_{4}F_{5}(\frac{19}{24},\frac{23}{24},\frac{31}{24},\frac{35}{24};\frac{3}{4},\frac{7}{8},\frac{5}{4},\frac{11}{8},\frac{3}{2};-\frac{729}{262144z^6})
\end{math}\\
& \begin{math}
-\frac{63}{4096}\frac{\sqrt{2}}{z^{\frac{9}{2}}\sqrt{\pi}}
{}_{4}F_{5}(\frac{11}{12},\frac{13}{12},\frac{17}{12},\frac{19}{12};\frac{7}{8},\frac{9}{8},\frac{11}{8},\frac{3}{2},\frac{13}{8};-\frac{729}{262144z^6})
\end{math}\\
& \begin{math}
+\frac{1989}{262144}\frac{\sqrt{2}\sin(\frac{3\pi}{8})}{z^{\frac{21}{4}}\Gamma(\frac{3}{4})}
{}_{4}F_{5}(\frac{25}{24},\frac{29}{24},\frac{37}{24},\frac{41}{24};\frac{9}{8},\frac{5}{4},\frac{3}{2},\frac{13}{8},\frac{7}{4};-\frac{729}{262144z^6})
\end{math}\\
\hline
$\frac{4}{5}$ & \begin{math}
\frac{16}{25}\frac{\Gamma(\frac{4}{5})\sin(\frac{2\pi}{5})}{z^{\frac{4}{5}}\pi}
{}_{6}F_{7}(\frac{9}{40},\frac{7}{20},\frac{19}{40},\frac{29}{40},\frac{17}{20},\frac{39}{40};
\frac{1}{5},\frac{3}{10},\frac{2}{5},\frac{1}{2},\frac{7}{10},\frac{4}{5},\frac{9}{10};
\frac{16384}{9765625 z^8})
\end{math}\\
& \begin{math}
-\frac{48}{125}\frac{\Gamma(\frac{3}{5})\sin(\frac{\pi}{5})}{z^{\frac{8}{5}}\pi}
{}_{6}F_{7}(\frac{13}{40},\frac{9}{20},\frac{23}{40},\frac{33}{40},\frac{19}{20},\frac{43}{40};
\frac{3}{10},\frac{2}{5},\frac{1}{2},\frac{3}{5},\frac{4}{5},\frac{9}{10},\frac{11}{10};
\frac{16384}{9765625 z^8})
\end{math}\\
& \begin{math}
-\frac{112}{625}\frac{csc(\frac{2\pi}{5})\sin(\frac{\pi}{5})}{z^{\frac{12}{5}}\Gamma(\frac{3}{5})}
{}_{6}F_{7}(\frac{17}{40},\frac{11}{20},\frac{27}{40},\frac{37}{40},\frac{21}{20},\frac{47}{40};
\frac{2}{5},\frac{1}{2},\frac{3}{5},\frac{7}{10},\frac{9}{10},\frac{11}{10},\frac{6}{5};
\frac{16384}{9765625 z^8})
\end{math}\\
& \begin{math}
+\frac{176}{3125}\frac{csc(\frac{\pi}{5})\sin(\frac{2\pi}{5})}{z^{\frac{16}{5}}\Gamma(\frac{4}{5})}
{}_{6}F_{7}(\frac{21}{40},\frac{13}{20},\frac{31}{40},\frac{41}{40},\frac{23}{20},\frac{51}{40};
\frac{1}{2},\frac{3}{5},\frac{7}{10},\frac{4}{5},\frac{11}{10},\frac{6}{5},\frac{13}{10};
\frac{16384}{9765625 z^8})
\end{math}\\
& \begin{math}
-\frac{6384}{78125}\frac{\Gamma(\frac{4}{5})\sin(\frac{2\pi}{5})}{z^{\frac{24}{5}}\pi}
{}_{6}F_{7}(\frac{29}{40},\frac{17}{20},\frac{39}{40},\frac{49}{40},\frac{27}{20},\frac{59}{40};
\frac{7}{10},\frac{4}{5},\frac{9}{10},\frac{6}{5},\frac{13}{10},\frac{7}{5},\frac{3}{2};
\frac{16384}{9765625 z^8})
\end{math}\\
& \begin{math}
\frac{14352}{390625}\frac{\Gamma(\frac{3}{5})\sin(\frac{\pi}{5})}{z^{\frac{28}{5}}\pi}
{}_{6}F_{7}(\frac{33}{40},\frac{19}{20},\frac{43}{40},\frac{53}{40},\frac{29}{20},\frac{63}{40};
\frac{4}{5},\frac{9}{10},\frac{11}{10},\frac{13}{10},\frac{7}{5},\frac{3}{2},\frac{8}{5};
\frac{16384}{9765625 z^8})
\end{math}\\
& \begin{math} 
+\frac{26928}{1953125}\frac{csc(\frac{2\pi}{5})\sin(\frac{\pi}{5})}{z^{\frac{32}{5}}\Gamma(\frac{3}{5})}
{}_{6}F_{7}(\frac{37}{40},\frac{21}{20},\frac{47}{40},\frac{57}{40},\frac{31}{20},\frac{67}{40};
\frac{9}{10},\frac{11}{10},\frac{6}{5},\frac{7}{5},\frac{3}{2},\frac{8}{5},\frac{17}{10};
\frac{16384}{9765625 z^8})
\end{math}\\
& \begin{math}
-\frac{35464}{9765625}\frac{csc(\frac{\pi}{5})\sin(\frac{2\pi}{5})}{z^{\frac{36}{5}}\Gamma(\frac{4}{5})}
{}_{6}F_{7}(\frac{41}{40},\frac{23}{20},\frac{51}{40},\frac{61}{40},\frac{33}{20},\frac{71}{40};
\frac{11}{10},\frac{6}{5},\frac{13}{10},\frac{3}{2},\frac{8}{5},\frac{17}{10},\frac{9}{5};
\frac{16384}{9765625 z^8})
\end{math}\\
\hline
\end{tabular}
}
\end{table}
\newpage
\begin{table}[!h]
\caption{The Fox's $H$-function for $1< \alpha \leq 2$.}
\centering
\label{tab:2}
{\small
\begin{tabular}{|c||l|} \hline
$\alpha$  & \hspace{3cm} $H(z)=\alpha z f(z)$ \\
\hline\hline
$\frac{4}{3}$     &  \begin{math}
\frac{6^{\frac{1}{4}}}{\sqrt{\pi}}z \Biggl( \frac{1}{2}\, \frac{csc(\frac{7\pi}{24})csc(\frac{11\pi}{24})}{csc(\frac{3\pi}{8})} \, \frac{\Gamma(\frac{19}{24})\Gamma(\frac{23}{24})\Gamma(\frac{5}{8})}{\Gamma(\frac{17}{24})\Gamma(\frac{13}{24})\Gamma(\frac{7}{8})} \,\,
{}_{4}F_{5}(\frac{7}{24},\frac{11}{24},\frac{19}{24},\frac{23}{24};\frac{1}{4},\frac{3}{8},\frac{1}{2},\frac{3}{4},\frac{7}{8};\frac{729 z^8}{262144})
\end{math}\\
&  \begin{math}
-\frac{5\sqrt{6}}{384} \, z^2 \, \frac{csc(\frac{\pi}{24})csc(\frac{5\pi}{24})}{csc(\frac{\pi}{8})} \, \frac{\Gamma(\frac{13}{24})\Gamma(\frac{17}{24})\Gamma(\frac{7}{8})}{\Gamma(\frac{23}{24})\Gamma(\frac{19}{24})\Gamma(\frac{5}{8})} \,\,
{}_{4}F_{5}(\frac{13}{24},\frac{17}{24},\frac{25}{24},\frac{29}{24};\frac{1}{2},\frac{5}{8},\frac{3}{4},\frac{9}{8},\frac{5}{4};\frac{729 z^8}{262144})
\end{math}\\
&  \begin{math}
+\frac{77}{1024} \, z^4 \, \frac{csc(\frac{7\pi}{24})csc(\frac{11\pi}{24})}{csc(\frac{3\pi}{8})}\, \frac{\Gamma(\frac{19}{24})\Gamma(\frac{23}{24})\Gamma(\frac{5}{8})}{\Gamma(\frac{17}{24})\Gamma(\frac{13}{24})\Gamma(\frac{7}{8})} \,\,
{}_{4}F_{5}(\frac{19}{24},\frac{23}{24},\frac{31}{24},\frac{35}{24};\frac{3}{4},\frac{7}{8},\frac{5}{4},\frac{11}{8},\frac{3}{2};\frac{729 z^8}{262144})
\end{math}\\
&  \begin{math}
-\frac{221\sqrt{6}}{196608}\, z^6 \, \frac{csc(\frac{\pi}{24})csc(\frac{5\pi}{24})}{csc(\frac{\pi}{8})} \, \frac{\Gamma(\frac{13}{24})\Gamma(\frac{17}{24})\Gamma(\frac{7}{8})}{\Gamma(\frac{23}{24})\Gamma(\frac{19}{24})\Gamma(\frac{5}{8})} \,\,
{}_{4}F_{5}(\frac{25}{24},\frac{29}{24},\frac{37}{24},\frac{41}{24};\frac{9}{8},\frac{5}{4},\frac{3}{2},\frac{13}{8},\frac{7}{4};\frac{729 z^8}{262144})\Biggr)
\end{math}\\
\hline
$\frac{5}{4}$     &  \begin{math}
\frac{2^{\frac{7}{5}}}{\sqrt{5\pi}}z \Biggl(\frac{ csc(\frac{9\pi}{40})csc(\frac{7\pi}{20})csc(\frac{19\pi}{40})}{csc(\frac{\pi}{5})csc(\frac{3\pi}{10})csc(\frac{2\pi}{5})} \, \frac{\Gamma(\frac{29}{40})\Gamma(\frac{17}{20})\Gamma(\frac{39}{40})\Gamma(\frac{3}{5})}{\Gamma(\frac{9}{10})\Gamma(\frac{31}{40})\Gamma(\frac{13}{20})\Gamma(\frac{21}{40})} \,\,
{}_{6}F_{7}(\frac{9}{40},\frac{7}{20},\frac{19}{40},\frac{29}{40},\frac{17}{20},\frac{39}{40};\frac{1}{5},\frac{3}{10},\frac{2}{5},\frac{1}{2},\frac{7}{10},\frac{4}{5},\frac{9}{10};-\frac{16384 z^{10}}{9765625})
\end{math}\\
&  \begin{math}
-\frac{7 2^{\frac{4}{5}}}{100} z^2 \frac{ csc(\frac{17\pi}{40})csc(\frac{\pi}{20})csc(\frac{7\pi}{40})}{csc(\frac{2\pi}{5})csc(\frac{\pi}{10})csc(\frac{\pi}{5})}\frac{\Gamma(\frac{11}{20})\Gamma(\frac{27}{40})\Gamma(\frac{37}{40})\Gamma(\frac{4}{5})}{\Gamma(\frac{23}{40})\Gamma(\frac{19}{20})\Gamma(\frac{33}{40})\Gamma(\frac{7}{10})}
{}_{6}F_{7}(\frac{17}{40},\frac{11}{20},\frac{27}{40},\frac{37}{40},\frac{21}{20},\frac{47}{40};\frac{2}{5},\frac{1}{2},\frac{3}{5},\frac{7}{10},\frac{9}{10},\frac{11}{10},\frac{6}{5};-\frac{16384 z^{10}}{9765625})
\end{math}\\
&  \begin{math}
+\frac{1}{2^{\frac{9}{10}}} z^4 \frac{csc(\frac{\pi}{8})csc(\frac{3\pi}{8})}{csc(\frac{\pi}{10})csc(\frac{\pi}{5})csc(\frac{3\pi}{5})csc(\frac{2\pi}{5})}
{}_{7}F_{8}(\frac{5}{8},\frac{3}{4},\frac{7}{8},1,\frac{9}{8},\frac{5}{4},\frac{11}{8};\frac{3}{5},\frac{7}{10},\frac{4}{5},\frac{9}{10},\frac{11}{10},\frac{6}{5},\frac{13}{10},\frac{7}{5};-\frac{16384 z^{10}}{9765625})
\end{math}\\
&  \begin{math}
-\frac{2^{\frac{2}{5}}897}{31250} z^6 \frac{csc(\frac{3\pi}{40})csc(\frac{13\pi}{40})csc(\frac{9\pi}{20})}{csc(\frac{\pi}{10})csc(\frac{3\pi}{10})csc(\frac{2\pi}{5})} \frac{\Gamma(\frac{33}{40})\Gamma(\frac{19}{20})\Gamma(\frac{23}{40})\Gamma(\frac{7}{10})}{\Gamma(\frac{37}{40})\Gamma(\frac{27}{40})\Gamma(\frac{11}{20})\Gamma(\frac{4}{5})}
{}_{6}F_{7}(\frac{33}{40},\frac{19}{20},\frac{43}{40},\frac{53}{40},\frac{29}{20},\frac{63}{40};\frac{4}{5},\frac{9}{10},\frac{11}{10},\frac{13}{10},\frac{7}{5},\frac{3}{2},\frac{8}{5};-\frac{16384 z^{10}}{9765625})
\end{math}\\
&  \begin{math}
+\frac{2^{\frac{1}{5}}4433}{3125000} z^8 \frac{csc(\frac{\pi}{40})csc(\frac{3\pi}{20})csc(\frac{11\pi}{40})}{csc(\frac{\pi}{10})csc(\frac{\pi}{5})csc(\frac{3\pi}{10})}\frac{\Gamma(\frac{21}{40})\Gamma(\frac{13}{20})\Gamma(\frac{31}{40})\Gamma(\frac{9}{10})}{\Gamma(\frac{39}{40})\Gamma(\frac{17}{20})\Gamma(\frac{29}{40})\Gamma(\frac{3}{5})}
{}_{6}F_{7}(\frac{41}{40},\frac{23}{20},\frac{51}{40},\frac{61}{40},\frac{33}{20},\frac{71}{40};\frac{11}{10},\frac{6}{5},\frac{13}{10},\frac{3}{2},\frac{8}{5},\frac{17}{10},\frac{9}{5};-\frac{16384 z^{10}}{9765625})\Biggr)
\end{math}\\
\hline
$\frac{6}{5}$     &  \begin{math}
\frac{2^{\frac{5}{6}}5^{\frac{1}{3}}}{\sqrt{\pi}}z \Biggl(\frac{1}{8}\frac{ csc(\frac{11\pi}{60})csc(\frac{17\pi}{60})csc(\frac{23\pi}{60})csc(\frac{29\pi}{60})}{csc(\frac{5\pi}{12})}\frac{\Gamma(\frac{41}{60})\Gamma(\frac{47}{60})\Gamma(\frac{53}{60})\Gamma(\frac{59}{60})\Gamma(\frac{7}{12})}{\Gamma(\frac{49}{60})\Gamma(\frac{43}{60})\Gamma(\frac{37}{60})\Gamma(\frac{31}{60})\Gamma(\frac{11}{12}}) \times
\end{math}\\
&  \begin{math}
\times {}_{8}F_{9}(\frac{11}{60},\frac{17}{60},\frac{23}{60},\frac{29}{60},\frac{41}{60},\frac{47}{60},\frac{53}{60},\frac{59}{60};\frac{1}{6},\frac{1}{4},\frac{1}{3},\frac{5}{12},\frac{1}{2},\frac{2}{3},\frac{3}{4},\frac{5}{6},\frac{11}{12};\frac{9765625 z^{12}}{8707129344})
\end{math}\\
&  \begin{math}
-\frac{5^{\frac{2}{3}}2^{\frac{1}{6}}3}{320} z^2 \frac{ csc(\frac{7\pi}{20})csc(\frac{9\pi}{20})csc(\frac{\pi}{20})csc(\frac{3\pi}{20})}{csc(\frac{5\pi}{12})csc(\frac{\pi}{12})}
{}_{8}F_{9}(\frac{7}{20},\frac{9}{20},\frac{11}{20},\frac{13}{20},\frac{17}{20},\frac{19}{20},\frac{21}{20},\frac{23}{20};\frac{1}{3},\frac{5}{12},\frac{1}{2},\frac{7}{12},\frac{2}{3},\frac{5}{6},\frac{11}{12},\frac{13}{12},\frac{7}{6};\frac{9765625 z^{12}}{8707129344})
\end{math}\\
&  \begin{math}
+\frac{10^{\frac{1}{3}}1729}{2488320 }z^4 \frac{ csc(\frac{\pi}{60})csc(\frac{7\pi}{60})csc(\frac{13\pi}{60})csc(\frac{19\pi}{60})}{csc(\frac{\pi}{12})}\frac{\Gamma(\frac{31}{60})\Gamma(\frac{37}{60})\Gamma(\frac{43}{60})\Gamma(\frac{49}{60})\Gamma(\frac{11}{12})}{\Gamma(\frac{59}{60})\Gamma(\frac{53}{60})\Gamma(\frac{47}{60})\Gamma(\frac{41}{60})\Gamma(\frac{7}{12}})\times
\end{math}\\
&  \begin{math}
\times
{}_{8}F_{9}(\frac{31}{60},\frac{37}{60},\frac{43}{60},\frac{49}{60},\frac{61}{60},\frac{67}{60},\frac{73}{60},\frac{79}{60};\frac{1}{2},\frac{7}{12},\frac{2}{3},\frac{3}{4},\frac{5}{6},\frac{13}{12},\frac{7}{6},\frac{5}{4},\frac{4}{3};\frac{9765625 z^{12}}{8707129344})
\end{math}\\
&  \begin{math}
-\frac{124729}{8957952}z^6 \frac{ csc(\frac{11\pi}{60})csc(\frac{17\pi}{60})csc(\frac{23\pi}{60})csc(\frac{29\pi}{60})}{csc(\frac{5\pi}{12})}\frac{\Gamma(\frac{41}{60})\Gamma(\frac{47}{60})\Gamma(\frac{53}{60})\Gamma(\frac{59}{60})\Gamma(\frac{7}{12})}{\Gamma(\frac{49}{60})\Gamma(\frac{43}{60})\Gamma(\frac{37}{60})\Gamma(\frac{31}{60})\Gamma(\frac{11}{12}})\times
\end{math}\\
&  \begin{math}
\times
{}_{8}F_{9}(\frac{41}{60},\frac{47}{60},\frac{53}{60},\frac{59}{60},\frac{71}{60},\frac{77}{60},\frac{83}{60},\frac{89}{60};\frac{2}{3},\frac{3}{4},\frac{5}{6},\frac{11}{12},\frac{7}{6},\frac{5}{4},\frac{4}{3},\frac{17}{12},\frac{3}{2};\frac{9765625 z^{12}}{8707129344})
\end{math}\\
&  \begin{math}
+\frac{5^{\frac{2}{3}}2^{\frac{1}{6}}429}{655360} z^8 \frac{ csc(\frac{\pi}{20})csc(\frac{3\pi}{20})csc(\frac{7\pi}{20})csc(\frac{9\pi}{20})}{csc(\frac{\pi}{12})csc(\frac{5\pi}{12})}
{}_{8}F_{9}(\frac{17}{20},\frac{19}{20},\frac{21}{20},\frac{23}{20},\frac{27}{20},\frac{29}{20},\frac{31}{20},\frac{33}{20};\frac{5}{6},\frac{11}{12},\frac{13}{12},\frac{7}{6},\frac{4}{3},\frac{17}{12},\frac{3}{2},\frac{19}{12},\frac{5}{3};\frac{9765625 z^{12}}{8707129344})
\end{math}\\
&  \begin{math}
-\frac{10^{\frac{1}{3}}596932063}{16717688340480}z^{10}\frac{ csc(\frac{\pi}{60})csc(\frac{7\pi}{60})csc(\frac{13\pi}{60})csc(\frac{19\pi}{60})}{csc(\frac{\pi}{12})}\frac{\Gamma(\frac{31}{60})\Gamma(\frac{37}{60})\Gamma(\frac{43}{60})\Gamma(\frac{49}{60})\Gamma(\frac{11}{12})}{\Gamma(\frac{59}{60})\Gamma(\frac{53}{60})\Gamma(\frac{47}{60})\Gamma(\frac{41}{60})\Gamma(\frac{7}{12}})\times
\end{math}\\
&  \begin{math}
\times
{}_{8}F_{9}(\frac{61}{60},\frac{67}{60},\frac{73}{60},\frac{79}{60},\frac{91}{60},\frac{97}{60},\frac{103}{60},\frac{109}{60};\frac{13}{12},\frac{7}{6},\frac{5}{4},\frac{4}{3},\frac{3}{2},\frac{19}{12},\frac{5}{3},\frac{7}{4},\frac{11}{6};\frac{9765625 z^{12}}{8707129344})\Biggr)
\end{math}\\
\hline
\end{tabular}
}
\end{table}
\newpage
\section{Conclusions}
\label{conclus}

In this article we presented analytic expressions for the $\alpha$-stable p.d.f.'s for rational values of the index $\alpha \in (0,2]$. These p.d.f.'s can be viewed as solutions of the spatial anomalous diffusion equation subjected to a Dirac delta initial condition. We established their connection to the Fox's $H$-function for the most general law $S_{\alpha}(\beta,K_{\alpha},\tau)$. The characteristic function of the $\alpha=1$ stable law was also reproduced by solving a suitably chosen fractional diffusion equation. An alternative way of classification which captures the infinite divisible character of stable laws was proposed.   The rationality of the index allows us to write closed expressions for the p.d.f.'s in terms of generalized hypergeometric functions. This method recovers known results, such as the Cauchy p.d.f. for $\alpha =1$, the Holtsmark p.d.f. for $\alpha={3 \over 2}$ and the normal p.d.f. for $\alpha=2$. When $\alpha$ takes an arbitrary rational value in $(0,2]$ new p.d.f.'s are derived generalizing and unifying previous results.
\addcontentsline{toc}{subsection}{Appendix A }
\section*{Appendix A }
\renewcommand{\theequation}{A.\arabic{equation}}
\setcounter{equation}{0}
The proof of (\ref{3a3}) is straightforward provided that we first show that
\begin{description}
\item[($i$)] If $f$ and $g$ belong to the space $AC^{[\alpha]}({\mathbb R})$
\footnote{This space consists of all functions $f$ which have continuous
derivatives up to order $[\alpha]-1$ on ${\mathbb R}$ with $f^{([\alpha]-1)}(x)\in
AC({\mathbb R})$. We also recall that $AC({\mathbb R})\subset \mathcal{S}({\mathbb R})$.} with boundary
conditions $\lim_{|x|\rightarrow \infty}f^{(k)}(x)=0$ $=\lim_{|x|\rightarrow
\infty}g^{(k)}(x)$, $k=0,\cdots, [\alpha]-1$ then
\begin{eqnarray}
\int_I \left({}_{-\infty}\mathcal{D}_x^{\alpha}f(x)\right)g(x)dx=
\int_I f(x)\, \left({}_{x}\mathcal{D}_{\infty}^{\alpha}g(x)\right) dx
\label{ia1}
\end{eqnarray}
\noindent where $I=(-\infty,\infty)$.
\par \textit{Proof of (\ref{ia1})}
\begin{eqnarray}
\int_I \left({}_{-\infty}\mathcal{D}_x^{\alpha}f(x)\right)g(x)dx
&=& \frac{1}{\Gamma([\alpha]+1-\alpha)}\int_I \left(\frac{d^{[\alpha]+1}}{dx^{[\alpha]+1}}\int_{0}^{\infty} \frac{f(x-u)}{u^{\alpha-[\alpha]}}du\right)g(x)dx \nonumber \\
&=& \frac{(-1)^{[\alpha]+1}}{\Gamma([\alpha]+1-\alpha)}\int_I \left(\int_{0}^{\infty} \frac{f(x-u)}{u^{\alpha-[\alpha]}}du\right)\frac{d^{[\alpha]+1}}{dx^{[\alpha]+1}}g(x)dx \nonumber \\
&=& \frac{(-1)^{[\alpha]+1}}{\Gamma([\alpha]+1-\alpha)}\int_I f(s)\left(\frac{d^{[\alpha]+1}}{ds^{[\alpha]+1}}\int_{s}^{\infty}\frac{g(x)}{(x-s)^{\alpha-[\alpha]}}dx\right)ds \nonumber \\
&=&\int_I f(x)\, \left({}_{x}\mathcal{D}_{\infty}^{\alpha}g(x)\right)dx. 
\label{ia2}
\end{eqnarray}
\item[($ii$)] The two identities hold
\begin{eqnarray}
{}_{-\infty}\mathcal{D}_x^{\alpha}e^{ipx}&=& (ip)^{\alpha}e^{ipx}=|p|^{\alpha}e^{\frac{i\alpha \pi}{2} {\rm sign}(p)}e^{ipx}, \nonumber \\
{}_{x}\mathcal{D}_{\infty}^{\alpha}e^{ipx}&=&
(-ip)^{\alpha}e^{ipx}=|p|^{\alpha}e^{- \frac{i\alpha \pi}{2} {\rm
sign}(p)}e^{ipx}.
\label{ia3}
\end{eqnarray}
\par \textit{Proof of (\ref{ia3})}\\
Using (\ref{3a2}) we have
\begin{eqnarray}
{}_{-\infty}\mathcal{D}_x^{\alpha}e^{ipx}&=& \frac{1}{\Gamma([\alpha]+1-\alpha)}\left(\frac{d^{[\alpha]+1}}{dx^{[\alpha]+1}}e^{ipx}\right) \int_0^{\infty}\frac{e^{-ipu}}{u^{\alpha-[\alpha]}}du \nonumber \\
&=& (ip)^{\alpha}e^{ipx}, \quad p>0.
\label{ia4}
\end{eqnarray}
In the derivation of (\ref{ia4}) we have used the integral formulas
\cite{Ref25}
\begin{eqnarray}
\int_{0}^{\infty} x^{\mu -1}\cos(px) dx&=&\frac{\Gamma(\mu)}{p^{\mu}}\cos\left(\frac{ \mu \pi}{2}\right) \nonumber \\
\int_{0}^{\infty} x^{\mu -1}\sin(px) dx &=& \frac{\Gamma(\mu)}{p^{\mu}} \sin \left(\frac{\mu \pi}{2}\right), \quad \textrm{for}\,\, p>0\,\, \textrm{and} \,\, 0<|Re \mu|<1.
\label{ia5}
\end{eqnarray}
\end{description}
\addcontentsline{toc}{subsection}{Appendix B }
\section*{Appendix B }
\renewcommand{\theequation}{B.\arabic{equation}}
\setcounter{equation}{0}

We first prove the identities
\begin{description}
\item[($1$)]
\begin{eqnarray}
\lim_{\epsilon \rightarrow 0^+}\left(|x|^{\epsilon}-\frac{2}{\epsilon}\delta(x)\right)=\frac{1}{|x|}.
\label{B1}
\end{eqnarray}
\textit{Proof of (\ref{B1})}\\ Differentiating with respect to $x$ the expression
\begin{eqnarray}
\ln(|x|){\rm sign}(x)=\lim_{\epsilon \rightarrow 0^+}\left(\frac{|x|^{\epsilon}-1}{\epsilon}\right){\rm sign}(x) 
\label{B2}
\end{eqnarray}
and taking into account that $\frac{d {\rm sign}(x)}{dx}=2\delta(x)$ we
easily derive (\ref{B1}).
\item[($2$)]
\begin{eqnarray}
\mathcal{F}[|x|^{\alpha}]=\frac{2}{(2\pi|p|)^{1+\alpha}}\Gamma(1+\alpha)\cos\left(\frac{\pi}{2}(1+\alpha)\right), \quad \alpha>-1.
\label{B3}
\end{eqnarray}
\textit{Proof of (\ref{B3})}\\ This is a direct consequence of (\ref{ia5}).
\end{description}
The Fourier transform of $\ln(|x|)$ is
\begin{eqnarray}
\mathcal{F}[\ln(|x|)]&=&\lim_{\epsilon \rightarrow 0^+}\mathcal{F}\left(\frac{1-|x|^{-\epsilon}}{\epsilon}\right)=\lim_{\epsilon \rightarrow 0^+}\left(\frac{\delta(p)}{\epsilon}-\frac{2}{(2\pi|p|)^{1-\epsilon}}\Gamma(1-\epsilon)\frac{\sin\left(\frac{\pi}{2}\epsilon\right)}{\epsilon} \right) \nonumber \\
&=& -\frac{1}{2}\lim_{\epsilon \rightarrow 0^+}\left(|p|^{\epsilon -1}-2\frac{\delta(p)}{\epsilon} -C_{\gamma}\epsilon |p|^{\epsilon -1}\right) \nonumber \\
&=& -\frac{1}{2}\frac{1}{|p|}+C_{\gamma}\delta(p)
\label{B4}
\end{eqnarray}
where $C_{\gamma}$ is the Euler-Mascheroni constant. In the derivation of
(\ref{B4}) we made use of the $\Gamma(1+z), \, |z|<1$ \cite{Ref27} expansion
\begin{eqnarray}
\Gamma(1+z)=\sum_{n=0}^{\infty}a_n z^n
\label{B5}
\end{eqnarray}
where the coefficients are given by
\begin{eqnarray}
a_0=1, \quad na_n=-\gamma a_{n-1}+\sum_{k=2}^{n}(-1)^k a_{n-k}\zeta(k).
\label{B6}
\end{eqnarray}

\bibliographystyle{plain}

\begin{thebibliography} {3}

\bibitem{Ref1} L\'evy, P. (1925). Calcul des Probabilit\'es, Gauthier-Villars.

\bibitem{Ref2} Gnedenko, B. V., and Kolmogorov, A. N. (1968). Limit Distributions for Sums of Independent Random Variables, Addison-Wesley.

\bibitem{Ref3} Bardou, F., Bouchaud, J. P., Aspect, A., and Cohen-Tannoudji, C. (2002). L\'evy Statistics and Laser Cooling, Cambridge University Press, London.

\bibitem{Ref4} Shlesinger, M.F., West, B.J., and Klafter, J. (1987). L\'evy dynamics of enhanced diffusion: Application to turbulence. \prl{58}{1100}.

\bibitem{Ref5} Shlesinger, M.F., Zaslavsky, and G.M., Frisch, U. (eds.) (1995). L\'evy Flights and Related Topics in Physics, Springer-Verlag.

\bibitem{Ref6} Nikias, C.L., and Shao, M. (1995). Signal Processing with Alpha-Stable Distributions and Applications, John Wiley and Sons, New York.

\bibitem{Ref7} Levandowsky, M., White, B.S., and Schuster, F.L. (1997). Random movements of soil amebas. \apt{36}{237}.

\bibitem{Ref8} Mantegna, R.N., Stanley, and H.E. (1997. Econophysics: scaling and its breakdown, \jsp{89}{469}.

\bibitem{Ref9} F. Mainardi, Yu. Luchko and G. Pagnini (2001). The Fundamental Solution of the Space-Time Fractional Diffusion Equation \fcaa{Vol. 4, No 2}{153}.

\bibitem{Ref10} F. Mainardi, G Pagnini and R. K. Saxena (2005). Fox $H$ Functions in Fractional Diffusion \jcam{Vol 178, No 1-2}{321}.

\bibitem{Ref11} Feller, W. (1971). An Introduction to Probability Theory and Its
Application, Vol. II, 2nd ed., John Wiley and Sons, New York.

\bibitem{Ref12} Sato, Ken-Iti (2004). L\'evy Processes and Infinitely Divisible Distributions, Cambridge Studies in Advanced Mathematics 68, Cambridge Univeristy Press, United Kingdom.

\bibitem{Ref13} Applebaum, D. (2005). L\'evy Processes and Stochastic Calculus, Cambridge Studies in Advanced Mathematics 93, Cambridge Univeristy Press, United Kingdom.

\bibitem{Ref14} Ma, X., Nikias, C.L.: On Blind Channel Identification for Impulsive Signal Enviroments, in Proc. ICASSP'95 (Detroit, MI), May.

\bibitem{Ref15} Zolotarev, V.M. (1986). One-Dimensional Stable Distributions. Amer. Math. Soc., Providence, RI.

\bibitem{Ref16} Fox, C. (1961). The $G$ and $H$ functions as symmetrical Fourier kernels. \tams{98}{395}.

\bibitem{Ref17} Mathai, and A.M., Saxena, R.K. (1970). Lecture Notes in Mathematics 348,
Generalized Hypergeometric Functions with Applications in
Statistics and Physical Sciences, Springer-Verlag.

\bibitem{Ref18} Schneider, W.R. (1985). Lecture Notes in Mathematics 1250, Stochastic Processes-Mathematics and Physics II, Springer-Verlag.

\bibitem{Ref19} Schneider, W.R., and Wyss, W. (1989). Fractional diffusion and wave equations. \jmp{30}{134}.

\bibitem{Ref20} West, B.J., Grigolini, P., Metzler, R., and Nonnenmacher, T.F. (1997). Fractional diffusion and L\'evy stable processes. \pre{55}{99}.

\bibitem{Ref21} Jespersen, S., Metzler, R., and Fogedby, H.C. (1999). L\'evy flights in external force fields: Langevin and fractional Fokker-Planck equations and their solutions. \pre{59}{2736}.

\bibitem{Ref22} Metzler, R., and Klafter, J. (2000). The random walk's guide to anomalous diffusion: A fractional dynamics approach. \pr{339}{1}.

\bibitem{Ref23} Samko, S.G., Kilbas, A.A., and Marichev, O.I. (1993). Fractional Integrals and
Derivatives - Theory and Applications, Gordon and Breach, New York.

\bibitem{Ref24} Miller, K.S., and Ross, B. (1993). An Introduction to the Fractional Calculus and Fractional Differential Equations, John Wiley and Sons, New York.

\bibitem{Ref25} Gradshteyn, I.S. and Ryzhik, I.M. (1994). Table of Integrals, Series, and Products, Academic Press.

\bibitem{Ref26} Holtsmark, J. (1919): \"Uber die Verbreiterung von Spektrallinien. \apk{363}{577}.

\bibitem{Ref27} Chandrasekhar, S. (1943). Stochastic Problems in Physics and Astronomy. \rmp{15}{1}.

\bibitem{Ref28} Luke, (1969). The special functions and their approximations Vol. 1, Academic Press.
\end{thebibliography}

\end{document}